\documentclass[review,12pt]{elsarticle}

\usepackage{amssymb}
\usepackage{amsmath}

\usepackage[english]{babel}
\usepackage[utf8]{inputenc}
\usepackage{amsmath,amsfonts}
\usepackage{graphicx}
\usepackage[margin=2cm]{geometry}
\usepackage{comment}
\usepackage{setspace}
\usepackage{booktabs, multirow, siunitx,makecell}

\usepackage{afterpage} 
\usepackage{float}
\usepackage{subcaption}
\usepackage{lineno} 

\usepackage{gensymb}
\usepackage{color} 

\usepackage[normalem]{ulem}
\usepackage{xcolor}

\newcommand{\err}[1]{\textcolor{black}{\,$\pm$\,#1}}

\biboptions{square,comma,sort&compress}

\journal{International Journal of Hydrogen Energy}

\begin{document}

\begin{frontmatter}



\title{Tailored heat treatments to characterise the fracture resistance of critical weld regions in hydrogen transmission pipelines}


\author[OXFORD]{Dannisa R. Chalfoun}

\author[EPRI]{Jonathan Parker}

\author[EPRI]{Michael Gagliano}


\author[OXFORD]{Emilio Mart\'{\i}nez-Pa\~neda\corref{cor1}}
\ead{emilio.martinez-paneda@eng.ox.ac.uk}

\address[OXFORD]{Department of Engineering Science, University of Oxford, Oxford OX1 3PJ, UK}
\address[EPRI]{Electric Power Research Institute, 3420 Hillview Avenue, Palo Alto, CA 94304, USA}

\cortext[cor1]{Corresponding author.}

\begin{abstract}
A new protocol is presented to directly characterise the toughness of microstructural regions present within the weld heat-affected zone (HAZ), the most vulnerable location governing the structural integrity of hydrogen transport pipelines. Heat treatments are tailored to obtain bulk specimens that replicate predominantly ferritic-bainitic, bainitic, and martensitic microstructures present in the HAZ. These are applied to a range of pipeline steels to investigate the role of manufacturing era (vintage versus modern), chemical composition, and grade. The heat treatments successfully reproduce the hardness levels and microstructures observed in the HAZ of existing natural gas pipelines. Subsequently, fracture experiments are conducted in air and pure H$_2$ at 100 bar, revealing a reduced fracture resistance and higher hydrogen embrittlement susceptibility of the HAZ microstructures, with initiation toughness values as low as 32 MPa$\sqrt{\text{m}}$. The findings emphasise the need to adequately consider the influence of microstructure and hard, brittle zones within the HAZ. \\

\end{abstract}




\end{frontmatter}




\section{Introduction}
\label{sec:Intro}

Establishing a backbone of hydrogen transmission pipelines is essential to realise the important role that hydrogen is expected to play in decarbonisation \cite{telessy2024repurposing,sayani2025techno,guzzo2025hydrogen}. However, this is challenged by the dramatic degradation in ductility, fracture toughness, and fatigue crack growth resistance experienced by pipeline steels in the presence of hydrogen, a phenomenon known as \emph{hydrogen embrittlement} \cite{djukic2019synergistic,malheiros2022local,yu2024hydrogen,chen2024hydrogen}. Exposing pipeline steels to hydrogen gas results in H$_2$ dissociation at the steel surface, followed by hydrogen absorption, diffusion, and subsequent embrittlement \cite{ronevich2018fatigue,jemblie2024safe,Depraetere2024TheSteel,Ronevich2021Hydrogen-assistedHydrogen}. In addition, the susceptibility to hydrogen embrittlement increases with hydrogen partial pressure; however, the relatively low volumetric energy density of hydrogen entails the need to operate at high H$_2$ pressures \cite{tahan2022recent,mandal2024computational,zhang2024modeling}. Thus, it is imperative to understand, characterise, and predict the structural integrity of hydrogen transmission pipelines for relevant H$_2$ pressures.\\

\begin{figure}[H]
\centering
\includegraphics[width=\textwidth]{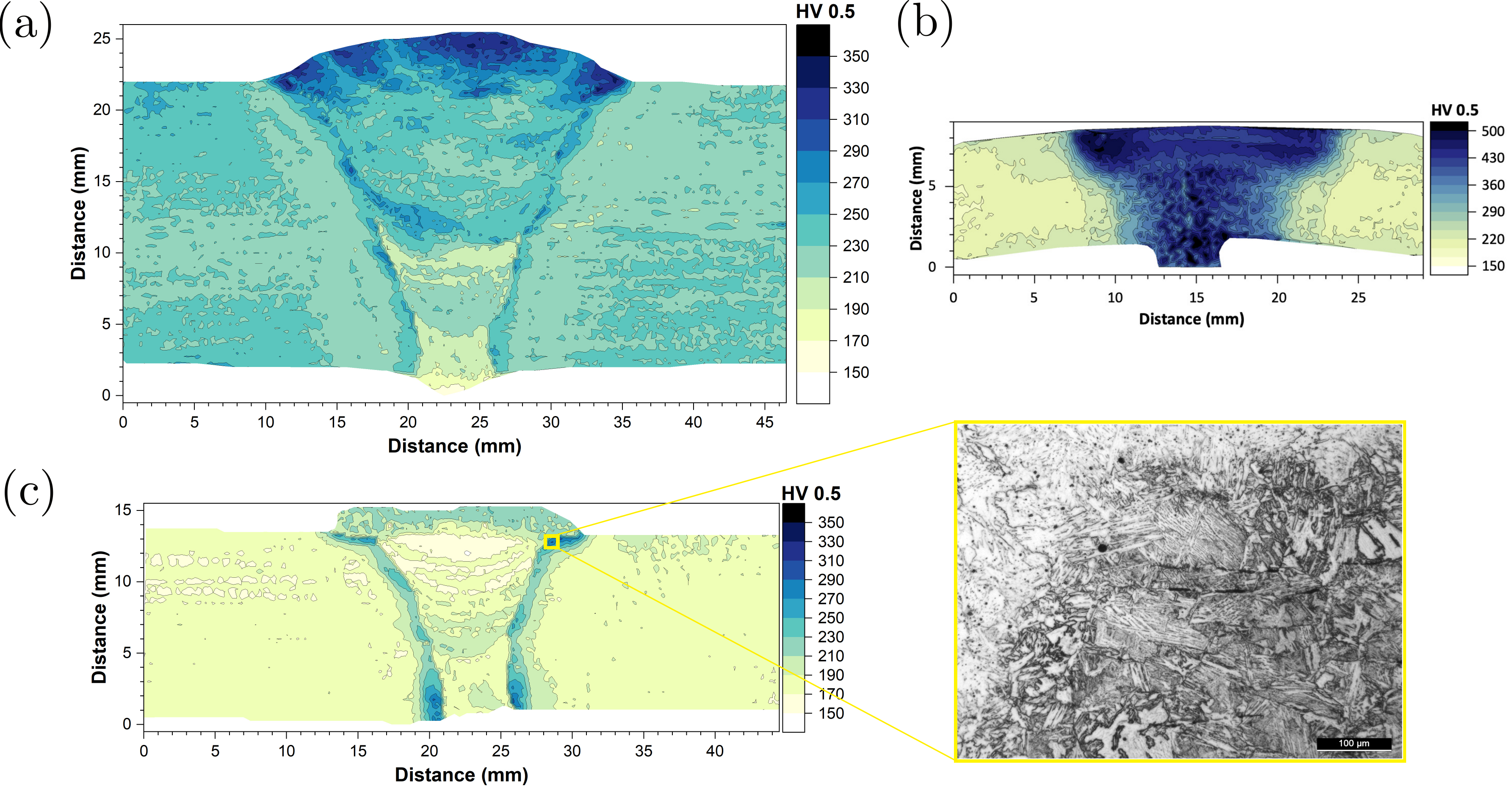}
\caption{Local hard, brittle regions within the weld compromise the integrity of hydrogen transmission pipelines. Hardness maps of welds associated with natural gas pipelines taken out of service: (a) submerged metal-arc welding (SMAW) girth weld of the X80 steel studied in this work, (b) electric resistance weld (ERW) seam weld of a vintage X46 steel, and (c) SMAW girth weld of the X60 vintage steel studied herein, with a corresponding optical micrograph confirming the presence of martensite in the HAZ. In all cases hardness values exceeding 350 HV (and, in some cases, 500 HV) are visible (vs base metal values in the range 150-200 HV).}
\label{fig:motivation_figure}
\end{figure}

Repurposing the existing natural gas pipeline network can be 5-10 times cheaper than constructing a new, dedicated hydrogen pipeline infrastructure and, consequently, governments and gas operators around the world have prioritised time-efficient and cost-effective retrofitting strategies \cite{brown2022development,vreeburg2025potential,mielich2025europe}. However, this brings additional structural integrity challenges. Existing assets contain a range of defects, which were introduced during original manufacturing and fabrication or have arisen through decades of operation and exposure to phenomena such as corrosion. The presence of defects results in local raises in stress and hydrogen concentration, thus promoting hydrogen-assisted cracking. Furthermore, natural gas pipelines comprise a wide variety of steel line pipes with varying age, grade, mechanical properties, and quality (carbon content, cleanliness), having been installed over a period exceeding five decades. To account for the evolution of steelmaking technologies and pipe manufacturing techniques, natural gas pipelines are typically classified into three categories: the \emph{early} era (pre-1940s), \emph{vintage} steels (1940s-1970s), and \emph{modern} (post-1970s). Vintage and modern steels are the most common, with the latter typically being cleaner (lower impurity concentration) and having a lower carbon content, a higher strength, and a higher fracture toughness \cite{Ronevich2024INFLUENCESTEELS,Kappes2023HydrogenMethods, Leis2015ManagingInfrastructure, API5L}. In terms of yield strength, vintage steels typically encompass grades between X42 and X60; i.e., with yield strength between $\sim$290 MPa (42 ksi) and $\sim$414 MPa (60 ksi), while modern steels include grades up to X100 ($\sim$690 MPa strength) and even X120 ($\sim$827 MPa strength). The material strength and hardness are of particular interest as stronger/harder alloys tend to exhibit a higher susceptibility to hydrogen embrittlement \cite{Gangloff2003,nanninga2010role}. Moreover, the presence of welds increases the heterogeneity of the pipeline system due to their associated gradient of non-equilibrium microstructures in the heat-affected zone (HAZ). Local heterogeneities in mechanical properties, hardness, fracture properties, and microstructure arise in the weld and, particularly, the HAZ. The main sources for these heterogeneities, include: (i) exposure of the base metal to a wide range of peak temperatures and rapid cooling rates that promote the formation of non-equilibrium phases in the HAZ; (ii) differences between the filler metal and the base metal chemical composition; (iii) subsequent re-heating due to multipass welding that can lead to the re-austenisation of the as-deposited material and the adjacent base metal; (iv) residual stresses; and (v) the presence of welding defects such as porosity, lack of fusion, and cracks. Hard spots and brittle phases, such as martensite, can be present in the HAZ of seam and girth pipeline welds, as shown in Fig.~\ref{fig:motivation_figure}. These features make welds the most vulnerable regions of the pipeline system and, as a result, significant efforts have been made to characterise the fracture and fatigue resistance of critical regions such as the HAZ \cite{Ronevich2021Hydrogen-assistedHydrogen}.\\

While the fracture and fatigue resistance of the base metal of natural gas pipelines has been extensively characterised in the presence of hydrogen (see, e.g., \cite{ronevich2018fatigue,jemblie2024safe,Depraetere2024TheSteel,Ronevich2021Hydrogen-assistedHydrogen} and Refs. therein), characterising the structural integrity of the critical weld regions remains an elusive challenge. This is despite the importance of quantifying the fracture resistance of susceptible weld regions, as these are the locations where cracking will initiate and therefore define safe operational conditions in the pipeline system. Furthermore, the fracture resistance of the base metal is likely to be very different to that of the harder and more brittle regions potentially present in the HAZ. The differences between weld regions can be observed in Fig. \ref{fig:motivation_figure}, where hardness maps and micrographs are provided for three pipeline seam welds removed from natural gas service. While the hardness values for the base metal regions are within the range of 150-200 HV, small regions of hardness exceeding 350 HV (and, on occasion, 500 HV) are observed in the HAZ. This 2-to 3-fold increase in hardness has notable implications for fracture resistance, given that hydrogen susceptibility increases significantly with material hardness \cite{Gangloff2003,nanninga2010role}. The very high hardness values observed also indicate the presence of susceptible microstructural phases (i.e., martensite), especially in vintage steels with high carbon content, as confirmed through optical microscopy in Fig. \ref{fig:motivation_figure}. Recently, Wijnen \textit{et al.} \cite{wijnen2025computational,wijnen2025} combined weld process modelling and phase field-based deformation-diffusion-fracture simulations to quantify the impact of these local hard, brittle zones. Their analysis found that failure pressures could be significantly lower than those defined through consideration of base material behaviour only. From an experimental viewpoint, characterisation of the properties of the HAZ and the critical regions within it is complex and hindered by their small size \cite{Alvaro2014}. As shown in Fig. \ref{fig:motivation_figure}, these local hard, brittle areas can span a few mm. While efforts are ongoing using miniature samples \cite{Madi2024MechanicalSpecimens} or through `ad hoc' tests that aim at placing a notch in the desired location \cite{Bortot2024InvestigationEnvironment}, these still measure a `composite' fracture toughness value that accounts for multiple microstructures (as opposed to the one of interest). In addition, quantitative local characterisation is further hindered by the weld geometry, existing residual stresses and local heterogenities, which can result in crack deflection, invalidating the test. The aim of this work is to circumvent these issues and enable direct, quantitative characterisation of the fracture resistance of the various regions present in a pipeline weld through appropriate heat treatments that simulate a weld thermal cycle to produce relevant HAZ microstructures in large samples amenable to fracture toughness testing.\\

In this work, we present a new protocol to quantitatively characterise, under exposure to relevant H$_2$ conditions, the fracture toughness of the microstructures typically contained in the HAZ of steel pipelines. Heat treatments are tailored to obtain bulk specimens containing a homogeneous microstructure with a given grain size and hardness according to the presence of non-equilibrium constituents. Coarse-grain bainitic and fine-grain ferritic-bainitic microstructures, as well as martensite, are evaluated for X60 and X80 steel pipes extracted from the natural gas grid. The effect of age and quality of the steel is also assessed, as the work considers both a vintage and a modern X60 steel. Mechanical properties and microstructures are characterised for the three pipeline steels considered and the four microstructural conditions (base metal and three tailored heat treatments). The resistance to crack initiation and growth is quantified through fracture experiments conducted in both air and H$_2$ environments (100 bar). The testing outputs from the simulated HAZ microstructures are compared with base metal data from this work and the literature, with results revealing that microstructural effects are significant and thus should be accounted for in the design and operation of hydrogen transmission pipelines.

\section{Experimental methods}
\label{sec:Methods}

\subsection{Materials}
Three pipeline steels were studied: (i) a \emph{vintage} X60 steel, referred to as X60V, (ii) a \emph{modern} X60, denoted X60M, and (iii) an X80 pipeline steel. The chemical composition of each steel is detailed in Table~\ref{tab:1}. These compositions are in line with the current American Petroleum Institute (API) 5L requirements. The resulting carbon equivalence (CE) values are also found to be within the required specifications \cite{API5L} (see the Supplementary Material).

\renewcommand{\arraystretch}{1.5} 
\begin{table}[H] 
    \centering
    \caption{Chemical composition of pipeline steels, wt.$\%$}
    \label{tab:1}
    \resizebox{\textwidth}{!}{ 
        \begin{tabular}{|c|c|c|c|c|c|c|c|c|c|c|c|c|c|c|c|} 
            \hline
            \textbf{Material} & \textbf{C} & \textbf{Si} & \textbf{Mn} & \textbf{P} & \textbf{S} & \textbf{Nb + V + Ti} & \textbf{Cr} & \textbf{Cu} & \textbf{Ni} & \textbf{Mo} & \textbf{Ca} & \textbf{Al} & \textbf{As} & \textbf{B} & \textbf{Fe}\\ \hline
            X60V & 0.178 & 0.219 & 1.30 & 0.016 & 0.043 & 0.096 & 0.028 & 0.018 & 0.042 & 0.009 & 0.0004 & 0.025 & 0.011 & 0.002 & 97.96\\ \hline
            X60M & 0.086 & 0.299 & 1.17 & 0.016 & 0.0017 & 0.045 & 0.025 & 0.009 & 0.021 & 0.0086 & 0.0024 & 0.019 & 0.0032 & \textless 0.0005 & 98.25\\ \hline
            X80 & 0.065 & 0.323 & 1.80 & 0.017 & \textless 0.001 & 0.070 & 0.202 & 0.191 & 0.195 & 0.014 & 0.0012 & 0.024 & 0.004 & \textless 0.0005 & 97.03\\ \hline
        \end{tabular}
    }
\end{table}

\begin{figure}[H]
\centering
\includegraphics[width=1\textwidth]{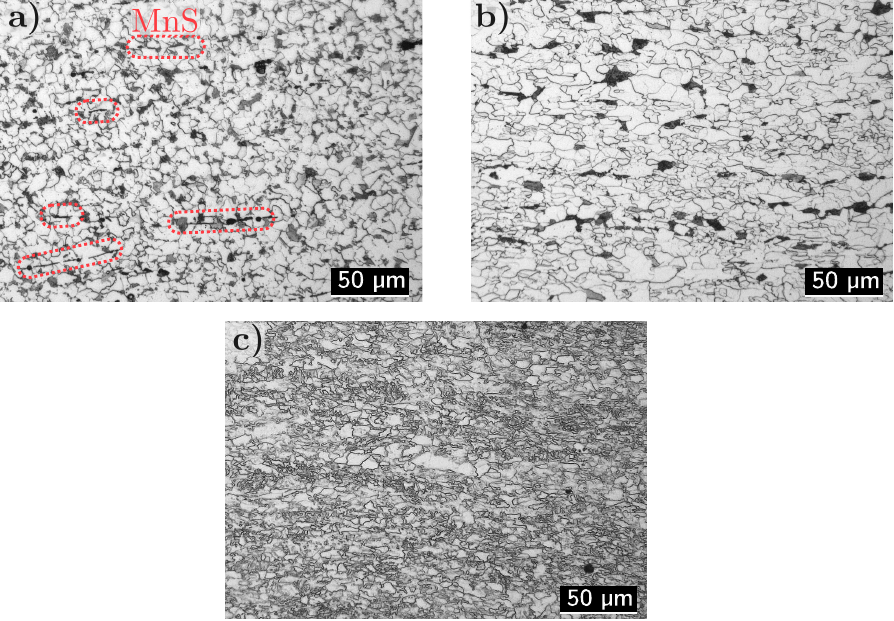}
\caption{Base metal microstructure for (a) the \emph{vintage} X60 steel (X60V), showing the presence of MnS inclusions (red dashed lines), (b) the \emph{modern} X60 steel (X60M), and (c) the X80 steel.}
\label{fig:bm_microstructures}
\end{figure}

The base metal microstructures, evaluated far from the welding regions, are illustrated in Fig.~\ref{fig:bm_microstructures}. These were examined using optical microscopy after grinding up to 1200 grit, polishing with diamond paste (6 and 1 $\mu$m), and etching with 2\% Nital. Both X60 steels exhibited a ferritic-pearlitic microstructure. Due to a higher carbon content, X60V presented a higher fraction of pearlite. Likewise, X60V also exhibited a higher density of non-metallic inclusions, such as manganese sulfides (MnS) stringers oriented parallel to the rolling direction. The base metal microstructure of the X80 steel showed acicular ferrite and bainite. The grain size of the base metal was measured using the intercept method, as described in ASTM (American Society for Testing and Materials) standard E112~\cite{ASTM2024ASTMSize}. This resulted in ASTM grain size 11 (7.1 \si{\micro\metre}) for both X60 steels and 12 (5.0 \si{\micro\metre}) for X80. Hardness, measured in the Vickers scale, was 175, 179, and 230 HV$_\text{10}$ for X60V, X60M, and X80, respectively.

\subsection{Illustrative weld characterisation}

While in this work tailored heat treatments are employed on base metal samples to mimic relevant weld microstructures (i.e., samples extracted from the weld region are not tested), optical micrographs are also extracted from the girth and seam welds of the X60V pipeline to illustrate the degree of microstructural heterogeneity present. Regardless of the age and the chemical composition variation, steel line pipes typically have a longitudinal seam weld and are joined by manual girth welds at their ends to form a pipeline~\cite{API5L}.\\
 
The type of welding process used and the welding procedure employed directly influence the microstructure and mechanical properties of the heat-affected zone (HAZ) and fusion zone (FZ). The factors that govern the resulting HAZ and FZ properties include the composition and microstructural condition of the base metal, heat input, pre-heat and interpass temperatures, the use and choice of filler metal, the number of weld passes, and the cooling rate. For example, seam welds typically have one or two passes and they are performed in a workshop during line pipe manufacturing using automated techniques such as Electric Resistance Welding (ERW), Flash Weld (although it was only used between 1930 and 1969), and Submerged-Arc Welding (SAW)~\cite{James2015FailurePipelines}. Girth welds are usually made in the field by qualified welders using a manual process such as Shielded Metal-Arc Welding (SMAW), giving place to a weldment with multiple passes depending on the size and thickness of the pipe ~\cite{James2015FailurePipelines}.\\

\begin{figure}[H]
\centering
\includegraphics[width=\textwidth]{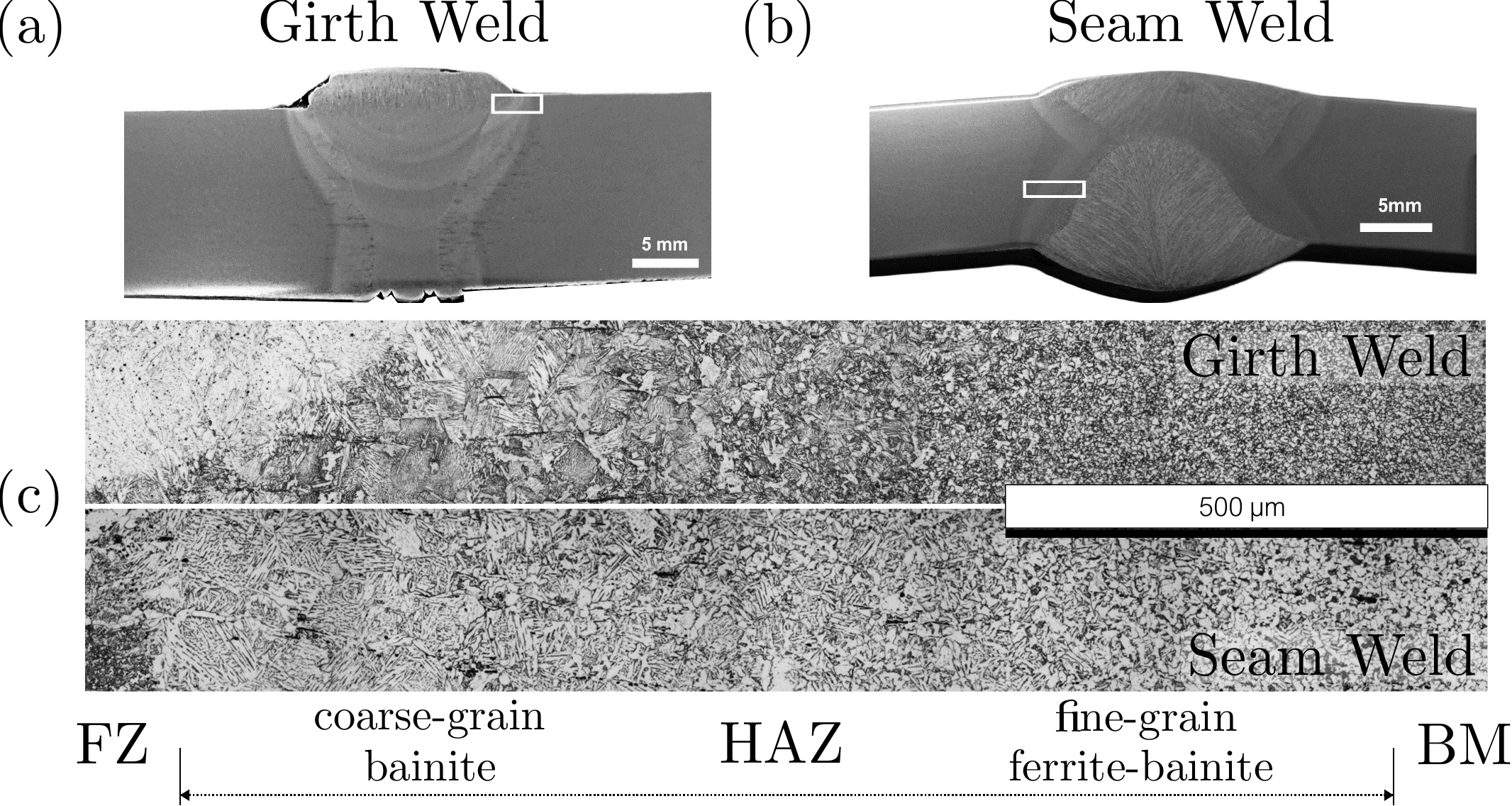}
\caption{Representative illustration of local weld heterogeneities: (a) Girth and (b) seam welds of the X60V pipeline characterised in this work. The optical micrographs in (c) illustrate the HAZ sub-zones between the fusion zone (FZ) and the base metal (BM), going from coarse-grain bainitic microstructures near the FZ to fine-grain ferritic-bainitic microstructures near the BM.}
\label{fig:girth_seam_weld}
\end{figure}

The fusion zone of pipeline steel welds is generally comprised of polygonal (proeutectoid) and acicular ferrite, although Widmanstätten side plates, pearlite, and upper bainite can also be present, if the cooling rate increases \cite{Stearling1992IntroductionWelding}. Conversely, the HAZ presents a gradient of non-equilibrium microstructures, which can be classified into different sub-zones. As illustrated in Fig.~\ref{fig:girth_seam_weld}, these sub-zones co-exist within a region extending over a few mm, resulting in a microstructural gradient across the HAZ and a corresponding variation in mechanical properties. 

The girth and seam welds of the X60V pipeline steel employed in this work are characterised to illustrate the HAZ microstructural variability, as shown in Fig. \ref{fig:girth_seam_weld}. Several HAZ sub-zones can be identified, with their specific characteristics depending on the thermal cycle (peak temperature, dwell time, cooling rate profile), the thermomechanical history of the parent steel, the chemical composition, and the stability of precipitates. A coarse-grain zone is present next to the fusion zone due to complete transformation to austenite, as local temperatures reach $\sim 1400~^\circ\text{C}$ during the thermal cycle. This zone remains at relatively high temperatures for a sufficient time for grains to grow, resulting in a coarse microstructure. 
Kinetic growth can be reduced due to the presence of undissolved carbides and/or nitrides, or enhanced if impurities such as P and S are segregated to the grain boundaries. Bainite is typically the dominant microstructure in this coarse-grain region. Steels with high CE can form hard phases, such as martensite, after rapid cooling. Steels with low CE typically present a higher fraction of proeutectoid ferrite. This region is prone to have the highest hardness and lowest fracture resistance  \cite{Bortot2024InvestigationEnvironment, Zerbst2014ReviewPerspective, Zafra2021FractureMicromechanisms}. Moving towards the base metal, a fine-grain microstructure is observed, as grain growth is restricted due to the maximum temperatures reached in this region (below $\sim 1100~^\circ\text{C}$) \cite{Stearling1992IntroductionWelding}. Upon cooling, ferrite can nucleate at the prior austenitic grain boundaries while carbon is segregated towards the centre of the grains, giving place to pearlite or bainite. Thus, ferrite and bainite dominate this fine-grain microstructure. Moving further towards the base metal, other sub-zones are identified, such as the partially transformed zone (where only pearlite transforms into austenite) and the spheroidized carbides zone, where Fe$_3$C spherical particles result from the spheroidization of the cementite present in the pearlite lamellae \cite {Stearling1992IntroductionWelding}. Of interest here is to characterise the fracture resistance of the bainitic and ferritic-bainitic microstructures that exist close to the fusion line, together with the martensite microstructures that can arise under certain conditions (as shown in Fig. \ref{fig:motivation_figure}).

\subsection{Heat treatments to simulate HAZ microstructures}

Base metal samples from each pipeline steel were extracted and subjected to heat treatments to simulate different microstructures that may be present in a weld HAZ. The samples consisted of cylindrical bars measuring 12.7 mm in diameter and 150 mm in length, machined parallel to the rolling direction of the pipes to obtain L-S fracture specimens, such that the tensile load is applied parallel to the longitudinal (rolling) direction, with crack propagation occurring along the short transverse direction. Three heat treatments were conducted on each of the pipeline steels considered (X60V, X60M, X80) to simulate the relevant microstructures shown in Figs. \ref{fig:motivation_figure} and \ref{fig:girth_seam_weld}. The first heat treatment, aimed at obtaining a fine-grain ferritic-bainitic microstructure (see Fig. \ref{fig:girth_seam_weld}), involved heating at $1000~^\circ\text{C}$ in a furnace for 300 s, followed by oil quenching (OQ). This treatment is denoted $1000~^\circ\text{C}$ OQ and, as intended and discussed in Section \ref{sec:Results}, resulted in a microstructure containing acicular ferrite and bainite, with hardness HV$_\text{10}$ values between 184 and 248, and grain sizes between 13 and 22 \si{\micro\metre}. The second heat treatment, labelled $1100~^\circ\text{C}$ OQ, involved heating at $1100~^\circ\text{C}$ for 300 seconds, followed by oil quenching (OQ). This heat treatment was aimed at replicating the coarse-grain bainitic microstructure zone (left region in Fig. \ref{fig:girth_seam_weld}). As discussed and illustrated in Section \ref{sec:Results}; a microstructure consisting primarily of upper bainite was obtained, with hardness values between 204 and 276 and grain sizes between 40 and 84.5 \si{\micro\metre}. Finally, the third heat treatment was aimed at replicating the hard, martensite microstructures shown in Fig. \ref{fig:motivation_figure}. To this end, samples were heat treatment at $1300~^\circ\text{C}$ for 10 seconds and water quenched (WQ) in a brine solution at $0~^\circ\text{C}$, a protocol denoted $1300~^\circ\text{C}$ WQ. This heat treatment effectively resulted in a martensite microstructure, as intended, with hardnesses between 338 and 440 HV and grain sizes spanning 46 to 130 \si{\micro\metre} (see Section \ref{sec:Results}). This corresponds to a plausible worst-case scenario of HAZ property behaviour. The heat treatment parameters, along with the targeted microstructures and main properties, are summarised in Table 2.\\

\begin{table}[H]
\centering
\caption{Heat treatment parameters and targeted properties.}
\label{tab:2}
\resizebox{\textwidth}{!}{
\begin{tabular}{lcccccc}
\toprule
\shortstack{\textbf{Heat}\\\textbf{Treatment}} & 
\shortstack{\textbf{Max. Temp.}\\\textbf{($^\circ$C)}} & 
\shortstack{\textbf{Holding}\\\textbf{Time (s)}} & 
\shortstack{\textbf{Cooling}\\\textbf{Media}} & 
\shortstack{\textbf{Targeted}\\\textbf{Microstructure}} & 
\shortstack{\textbf{Targeted}\\\textbf{Grain size} {\boldmath$(\si{\micro\metre})$}} & 
\shortstack{\textbf{Targeted}\\\textbf{Hardness} {\boldmath(HV$_{10}$)}} \\
\midrule
1000~$^\circ\text{C}$ OQ & 1000 & 300 & Oil           & Ferrite and Bainite & $>$10 & $>$180 \\
1100~$^\circ\text{C}$ OQ & 1100 & 300 & Oil           & Bainite             & $>$30 & $>$200 \\
1300~$^\circ\text{C}$ WQ & 1300 & 10  & Brine at 0 $^\circ$C & Martensite    & $>$40 & $>$300 \\
\bottomrule
\end{tabular}
}
\end{table}

\begin{figure}[H]
\centering
\includegraphics[width=\textwidth]{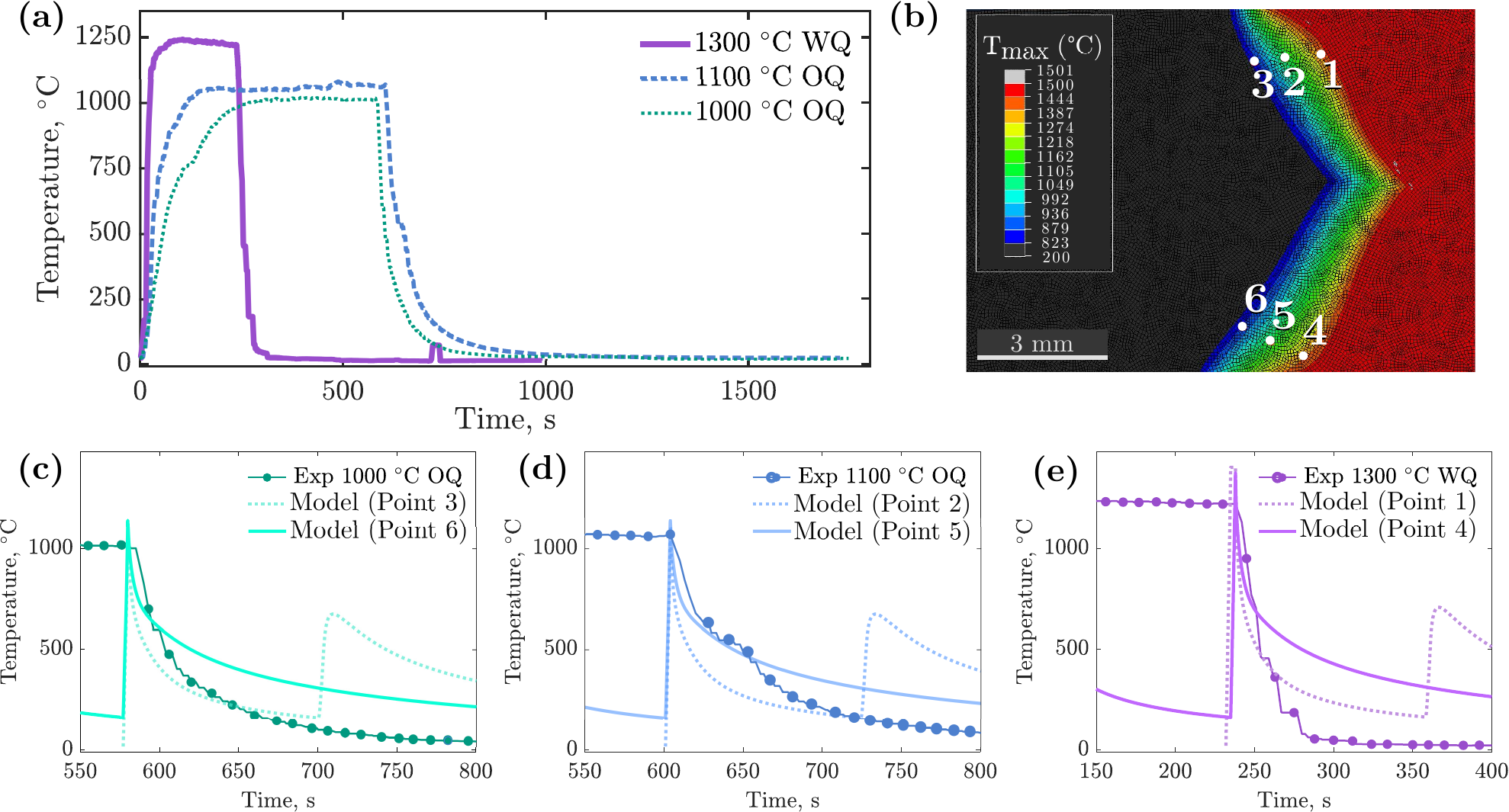}
\caption{Combining experiments and modelling to establish suitable heat treatments: (a) Temperature-time profile for the heat treatments conducted to replicate the HAZ microstructure in base metal cylinders; (b) Maximum temperature contour along the HAZ, obtained through a two-pass weld simulation using the model developed in Ref. \cite{wijnen2025computational}; and comparison of the experimental and predicted cooling curves for peak temperatures of (c) 1000~$^\circ\text{C}$ followed by oil quenching (OQ), (d) 1100~$^\circ\text{C}$ followed by oil quenching (OQ), and (e) 1300~$^\circ\text{C}$ followed by water quenching (WQ). The simulation points correspond to those highlighted in the $T_{max}$ contours. The second increase in temperature observed in the dashed curve for the first pass of the weld corresponds to the deposition of the second bead, shown by the solid line.}
\label{fig:peak_temp_cooling_rates}
\end{figure}

Thermo-metallurgical simulations were conducted using the model presented in Ref. \cite{wijnen2025computational} to assist in determining the appropriate heat treatments. As shown in Fig. \ref{fig:peak_temp_cooling_rates}, these were found to be in very good agreement with the temperature-time profiles (cooling rates) determined experimentally. The experimental profiles were obtained using a K-type thermocouple inserted at the centre of a dummy specimen with identical dimensions to the samples used for machining the fracture specimens. The model simulates a two-pass weld, with the top bead deposited first. Cooling curves corresponding to both weld passes were overlaid onto the experimental data to facilitate comparison of the cooling rate profiles. The heat treatments performed in this work replicate a single thermal cycle in which the peak temperature, dwell time, and cooling rate are varied to simulate the thermal profiles within specific regions of the weld HAZ and produce relevant microstructures. In the case of girth welds, multiple weld passes add further complexity to the resulting microstructure due to the overlap of thermal profiles, which can lead to grain refinement but also to the formation of local brittle zones \cite{Li2011InfluenceZones}. Nonetheless, the approach used herein provides the opportunity to characterise the mechanical and fracture properties of specific HAZ microstructures.

The mechanical properties of the resulting heat-treated specimens were assessed through hardness measurements and tensile tests. Hardness tests were conducted on a Vickers benchtop hardness meter using a 10 kg load (HV$_\text{10}$) and a dwell time of 15 s. Tensile tests were performed on a micromechanical testing machine using a displacement rate of 0.1 mm/s.

\subsection{Fracture toughness tests}

\begin{figure}[H]
\centering
\includegraphics[width=\textwidth]{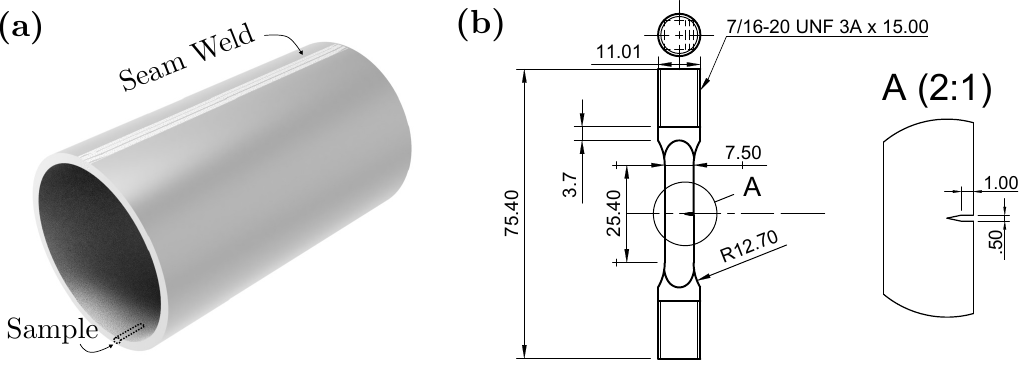}
\caption{Single-edge notched tension (SENT) specimen geometry and characteristics: (a) sample extraction location from the steel pipes, and (b) specimen layout (all dimensions in mm).}
\label{fig:SENT_spec}
\end{figure}

Single-edge notched tension (SENT) specimens were used to characterise the initiation and crack growth resistance of the three pipeline steels considered and the four processing conditions (three tailored heat treatments and the reference base metal). SENT samples provide triaxiality conditions that more closely resemble those of in-service pipelines, relative to other standardised testing configurations such as the Compact Tension (CT). The SENT specimens, with a square cross-section and threaded ends, were machined to produce a calibrated section with a width (W) and thickness (B) of 7.5 mm, a gauge length of 25.4 mm, and a total length of 75.4 mm, as illustrated in Fig.~\ref{fig:SENT_spec}b. The nominal thread diameter was 11.1 mm to ensure complete removal of the oxide layer and the decarburised region formed during the heat treatment in an unprotected atmosphere. In all specimens, a V-notch with a depth ($a_n$) of 1.0~mm and a width of 0.5~mm was introduced by electro-discharge machining (EDM) on the surface corresponding to the internal wall of the pipe, as shown in Fig.~\ref{fig:SENT_spec}b. The specimens were ground with SiC emery paper up to 1200 grit to obtain an average surface roughness equal to or below 0.8 µm, as recommended by the BS ISO 8571 standard for SENT specimens \cite{BSISO2018BSSpecimens}.\\ 

The specimens were pre-cracked under fatigue in air at a frequency of 10 Hz and with a load ratio of 0.1 to attain a pre-crack with nominal length $a_0$ = 0.19W = 1.4 mm. The growth of the pre-crack was monitored using the direct current potential drop (DCPD) technique, and the maximum load was kept equal to or below 40\% of the load required for plastic failure, as per ASTM E1820~\cite{E1820}. Once the pre-crack attained half of the target length, the load was reduced to maintain a constant stress intensity factor $K_I$, which was calculated using the expression for SENT specimens provided in Ref. \cite{Cupertino-Malheiros2024OnSusceptibility}. After pre-cracking, side grooves were machined by EDM to a nominal depth of 0.25 mm on each side (total of 7\%B). Side grooves can help to avoid crack deflection and tunnelling \cite{Verstraete2013DeterminationMeasurements, Hioe2017ComparingTesting}. All the specimens were re-ground with SiC emery paper up to 1200 grit before fracture testing to remove oxides formed during machining.\\  

The fracture tests were conducted in air and in a gaseous hydrogen environment, under an H$_2$ pressure of 100 bar (10 MPa). This pressure was taken as representative of the conditions relevant to hydrogen transmission in repurposed natural gas pipelines. For both the in-air and H$_2$ tests, a sufficiently slow displacement rate of 4 x 10$^{-5}$ mm/s was used to enable hydrogen redistribution to take place, maximising susceptibility. The Load-Line Displacement (LLD) was measured by means of an external Linear Variable Differential Transducer (LVDT), in both in-air and H$_2$ experiments. The LLD estimates were corrected to account for the machine compliance by testing a stiffer sample. The Direct Current Potential Drop (DCPD) method was used to monitor crack extension, in both the in-air and H$_2$ tests. Johnson's formula \cite{E1820} was used to convert the potential drop $V$ to crack length $a$, from the dimensions of the specimen and the half-distance between the DCPD wires ($y$);
\begin{equation}
a = \frac{2W}{\pi} \cos^{-1} \left[\frac{\cosh\left( \frac{\pi y}{2W} \right)}{\cosh \left(\frac{V}{V_0} \cosh^{-1} \left( \frac{\cosh\left( \frac{\pi y}{2W} \right) }{ \cos\left( \frac{\pi a_0}{2W} \right)}\right)\right)}\right]
\label{eq:5}
\end{equation}

\noindent Here, $V_0$ denotes the initial value of the potential drop, obtained after 1 h of stabilisation for the initial pre-crack $a_0$. To isolate the individual contributions to the DCPD measurement of plasticity and crack growth, the plasticity line was subtracted from the curve of potential drop versus the (corrected) LLD, as described in Refs. \cite{Harris2022,VanMinnebruggen2017CrackMeasurement,Ronevich2021Hydrogen-assistedHydrogen}. The crack extension was also measured after testing, with heat tinting being performed in a furnace for three hours at $200~^\circ\text{C}$.  The region corresponding to stable crack growth, $\Delta a$, was inspected by optical microscopy and determined using a five-point averaging procedure to resolve potential variations across the sample thickness. A robust quantification of crack extension as a function of the load enabled the determination of crack growth resistance curves ($J$-$R$ curves), following standardised protocols \cite{BSISO2018BSSpecimens,E1820}. Of interest here are the values of the $J$-integral at the initiation of crack growth, denoted $J_0$, and after 0.2 mm of crack propagation, denoted $J_{0.2}$ and estimated through the intersection of the $J$-$R$ curve with the blunting line \cite{E1820}. These values were converted to their corresponding $K$-values ($K_{J0}$, $K_{J0.2}$) through the usual relationship: $K_J = \sqrt{E J/(1-\nu^2)}$.\\

In the H$_2$ tests, a soaking time of 70 minutes was adopted before mechanical loading was applied, so as to accommodate the time required to reach 100 bar of H$_\text{2}$ pressure and to stabilise the DCPD signal. High-purity hydrogen (99.9999$\%$) was used during testing and high-purity nitrogen (99.999\%) was used as inert gas for purging. As recognised in the literature \cite{Somerday2013ElucidatingConcentrations}, it is of critical importance to reduce the presence of impurities. To this end, the hydrogen and impurity contents in the autoclave were estimated on the basis of an exhaustive purging procedure, ensuring an oxygen content well below 1 ppm.\\

For all the experiments in this work, at least two tests per condition were conducted, with results showing good replicability. More details of the fracture testing configuration and analysis are given in the Supplementary Material.

\section{Results and discussion}
\label{sec:Results}

\subsection{Heat treatments to replicate HAZ microstructures}

\subsubsection{Microstructure and hardness}

\begin{figure}[H]
\centering
\includegraphics[width=\textwidth]{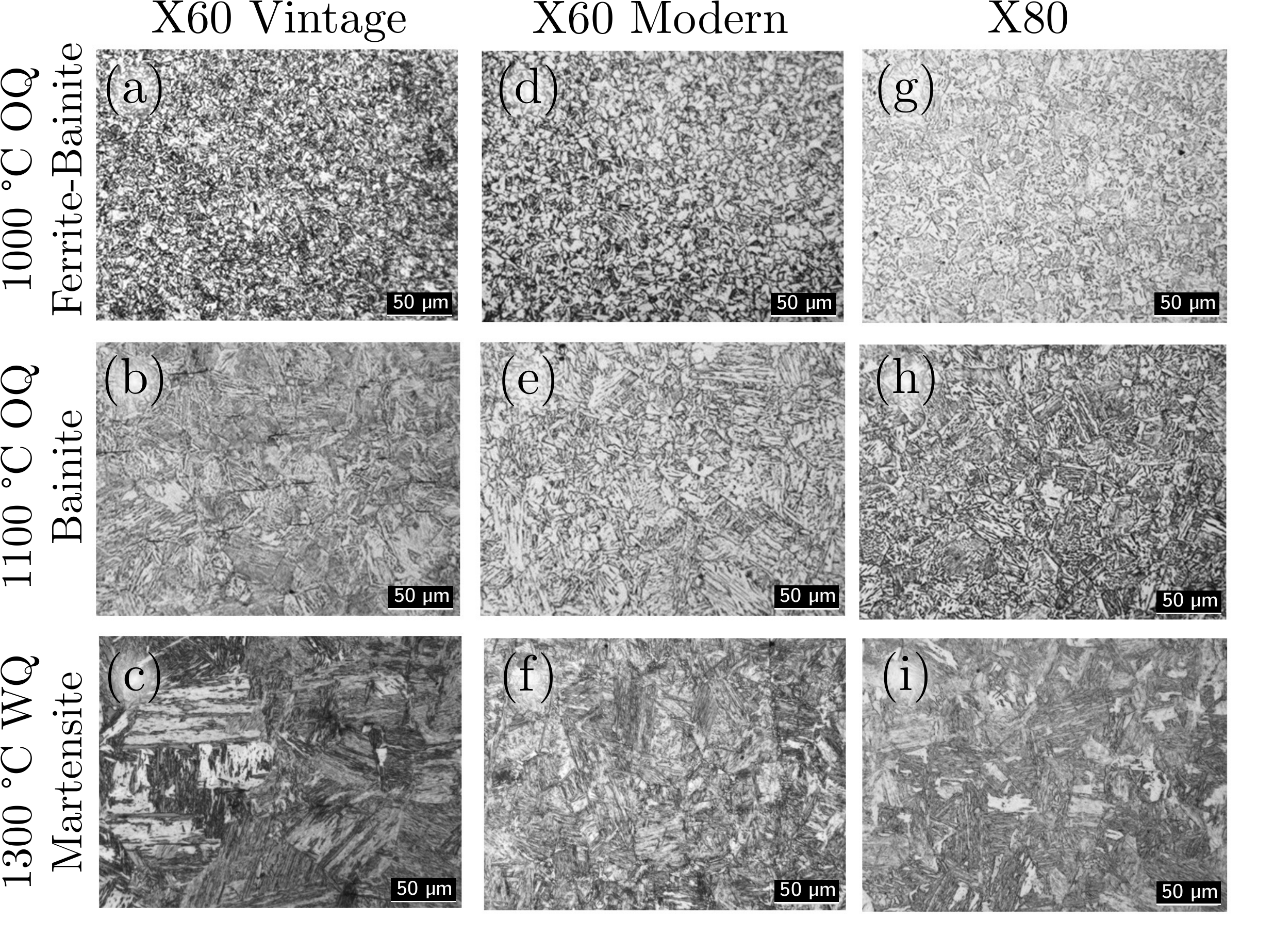}
\caption{Microstructures obtained after conducting tailored heat treatments on base metal samples of X60 vintage steel (left column), X60 modern steel (centre column) and X80 steel (right column). The heat treatments aimed at obtaining a fine-grain ferritic-bainitic microstructure (1000 $^\circ$C OQ, top row), a coarse-grain bainitic microstructure (1100 $^\circ$C OQ, centre row), and a martensitic microstructure (1300$^ \circ$C WQ, bottom row).}
\label{fig:microstructures_after_HT}
\end{figure}

A summary of the microstructures resulting from the heat treatments is provided in Fig. \ref{fig:microstructures_after_HT}, with quantitative mechanical and microstructural details being provided in Table \ref{tab:3}. After the heat treatments, the characteristic banding of the ferritic-pearlitic microstructure of the base metal (Fig. \ref{fig:bm_microstructures}) was no longer present in any of the steels studied. As intended, three characteristic microstructures are observed, with variations in their properties measured across steel grade and age.  

The steels treated at 1000 $^\circ$C and OQ exhibited a fine-grain microstructure, containing acicular ferrite and bainite - see Figs. \ref{fig:microstructures_after_HT}a, \ref{fig:microstructures_after_HT}d, and \ref{fig:microstructures_after_HT}g. As listed in Table \ref{tab:3}, hardness testing revealed values of 248, 184, and 240 HV$_\text{10}$ for X60V, X60M, and X80, respectively. The grain size was measured to be 13 \si{\micro\metre} for the X60V case and 22 \si{\micro\metre} for X60M and X80. This corresponds to roughly a threefold increase in grain size for the modern grades relative to their respective base metals. The coarse-grain microstructure obtained after the 1100 $^\circ$C OQ heat treatment consisted predominantly of upper bainite (see Figs. \ref{fig:microstructures_after_HT}b, \ref{fig:microstructures_after_HT}e, and \ref{fig:microstructures_after_HT}h). For X60M and X80 steels, the presence of acicular ferrite is also likely (see Figs. \ref{fig:microstructures_after_HT}e and \ref{fig:microstructures_after_HT}h). The resulting hardness values were 271, 204, and 276 HV$_\text{10}$ for X60V, X60M, and X80, respectively. The vintage steel exhibited significant grain growth in the coarse-grain condition, with an increase of approximately 6.5 times compared to its corresponding fine-grain condition. In contrast, the modern grades exhibited moderate grain growth, with increases below twofold compared to their respective fine-grain condition characteristics (see Table \ref{tab:3}). Finally, the steels treated at 1300 $^\circ$C and WQ presented a microstructure consisting almost entirely of martensite, with potentially a very small amount of retained austenite. In agreement with expectations, this microstructure showed the highest hardness values among all conditions studied: 440, 338, and 357 HV$_\text{10}$ for X60V, X60M, and X80 steel, respectively. The measured hardness values are comparable to the values reported on the hardness maps for ex-service natural gas pipeline welds provided in Fig. \ref{fig:motivation_figure}. This heat treatment also resulted in the largest prior austenite grain sizes; 130, 47, and 46 \si{\micro\metre} for (respectively) X60V, X60M, and X80 steel. The greatest grain growth was observed in the vintage X60 steel, whose size increased from 13 \si{\micro\metre} in the fine-grain ferritic-bainitic microstructure to 85 \si{\micro\metre} in the coarse-grain bainitic condition and to 130 \si{\micro\metre} in the martensitic microstructure. Conversely, the X60M steel presented similar grain sizes in both coarse-grain and martensitic conditions (44 - 47 \si{\micro\metre}). In the case of the X80 steel, further grain growth was evident in the martensitic condition, with a grain size of 46 \si{\micro\metre} compared to 31 \si{\micro\metre} in the coarse-grain bainitic microstructure. Overall, the results reveal a significant influence of the steel age. The vintage steel presented the largest grain size in both coarse-grain and martensitic microstructures, as well as the hardest martensite. This behaviour can be rationalised in terms of the chemical composition of the X60V. First, the lack of undissolved microalloy precipitates, such as Nb, Ti, and V carbonitrides, which hinder grain growth during a thermal cycle \cite{Stearling1992IntroductionWelding, Cochrane2013HAZSteels}, could have promoted austenite growth kinetics while heating at temperatures above 1000 $^\circ$C. Although the X60V steel presents Nb, Ti, and V as alloying elements, the manufacturing practices at the time may have been insufficient to effectively control the size, composition, and distribution of the carbonitrides \cite{Cochrane2013HAZSteels}, relative to the modern grades, which were obtained by thermomechanical processing (see Supplementary Material). Second, the higher hardnesses of the heat-treated microstructures obtained for X60V, compared to the modern steels exposed to the same heat treatments, are attributed to the higher carbon content of the vintage steel (0.176 wt.$\%$)~\cite{Kappes2023HydrogenMethods, Cochrane2013HAZSteels}.\\

The results show that the heat treatments chosen succeeded in simulating the microstructures and mechanical behaviour of the relevant HAZ sub-zones, with the microstructures presented in Fig. \ref{fig:microstructures_after_HT} being consistent with those typically found in the HAZ of actual, ex-service pipelines, as evidenced by the micrographs in Fig. \ref{fig:girth_seam_weld}. Li et al.  \cite{Li2011InfluenceZones} report hardness levels ranging between 195 and 215 HV for a modern X70 steel after exposure to a maximum temperature peak of 1000 $^\circ$C during a double thermal cycle using a Gleeble thermo-mechanical simulator. These values fall within the range observed for the fine-grain microstructures (1000 $^\circ$C and OQ) for the modern X60 (184 HV) and X80 (240 HV) steels investigated in this work. As a worst-case scenario, a nearly 100$\%$ martensitic microstructure was successfully produced to represent the conditions associated with the highest peak temperatures and rapid cooling rates experienced in the HAZ \cite{Zafra2021FractureMicromechanisms,Cochrane2013HAZSteels}. In fully martensitic structures, hardness increases with carbon content as a result of carbon dissolution in the austenite phase. The maximum hardness of a fully martensitic microstructure can be estimated in terms of the carbon content \cite{Nolan2005HardnessPipeline}. Such an estimate would result in the values of 448 HV, 369 HV, and 351 HV for X60V, X60M, and X80, respectively. The hardness levels obtained in this work for martensite-containing microstructures (Table~\ref{tab:3}) are in very good agreement with these values and with other results available from the literature. For example, for C-Mn steels with a carbon content of 0.17 wt.$\%$ and 0.07 wt.$\%$, hardness levels of 441 HV and 353 HV were respectively measured in the HAZ after the highest cooling rate, while simulating the thermal cycle of a weld \cite{Hart1991CompositionalSteels}.   

\begin{table}[H]
\centering
\caption{Mechanical and microstructural properties for the base metal and the samples aiming at replicating the HAZ subzone microstructures: ferrite-bainite (1000 $^\circ$C OQ), bainite (1100 $^\circ$C OQ), and martensite (1300 $^\circ$C WQ).}
\label{tab:3}
\resizebox{\textwidth}{!}{
\begin{tabular}{l ccc ccc ccc ccc}
\toprule
\textbf{Property} &
\multicolumn{3}{c}{\textbf{Base Metal}} &
\multicolumn{3}{c}{\makecell{\textbf{Ferrite--Bainite}\\ \textbf{1000~$^\circ$C OQ}}} &
\multicolumn{3}{c}{\makecell{\textbf{Bainite}\\ \textbf{1100~$^\circ$C OQ}}} &
\multicolumn{3}{c}{\makecell{\textbf{Martensite}\\ \textbf{1300~$^\circ$C WQ}}}
\\
\cmidrule(lr){2-4}\cmidrule(lr){5-7}\cmidrule(lr){8-10}\cmidrule(lr){11-13}
 & X60V & X60M & X80 & X60V & X60M & X80 & X60V & X60M & X80 & X60V & X60M & X80 \\
\midrule
\makecell{Yield strength\\(MPa)} 
 & 407\err{1} & 417\err{4} & 576\err{10}
 & 469\err{28} & 352\err{5} & 565\err{10}
 & 524\err{10} & 415\err{23} & 527\err{5}
 & 1070\err{41} & 822\err{10} & 925\err{15}
\\
\makecell{Ultimate tensile\\strength (MPa)}
 & 500\err{21} & 483\err{10} & 648\err{4}
 & 738\err{14} & 559\err{1} & 728\err{5}
 & 757\err{8} & 571\err{5} & 684\err{0}
 & 1265\err{28} & 1030\err{66} & 1069\err{14}
\\
\makecell{Strain at failure\\$\varepsilon_{\mathrm f}$}
 & 0.23\err{0.00} & 0.19\err{0.02} & 0.15\err{0.02}
 & 0.16\err{0.01} & 0.25\err{0.01} & 0.16\err{0.01}
 & 0.16\err{0.02} & 0.18\err{0.03} & 0.16\err{0.02}
 & 0.04\err{0.01} & 0.06\err{0.00} & 0.04\err{0.02}
\\
\makecell{Hardness\\(HV$_{10}$)}
 & 175\err{6} & 179\err{8} & 230\err{9}
 & 248\err{6} & 184\err{14} & 240\err{20}
 & 271\err{29} & 204\err{12} & 276\err{18}
 & 440\err{6} & 338\err{8} & 357\err{26}
\\
\makecell{Grain size\\(\si{\micro\metre})}
 & 10.1\err{3} & 7.6\err{3} & 7.5\err{2}
 & 13\err{4} & 22\err{4} & 22\err{6}
 & 85\err{25} & 44\err{11} & 31\err{9}
 & 130\err{25} & 47\err{11} & 46\err{14}
\\
\bottomrule
\end{tabular}%
}
\end{table}

\subsubsection{Tensile behaviour}

\begin{figure}[H]
\centering
\includegraphics[width=\textwidth]{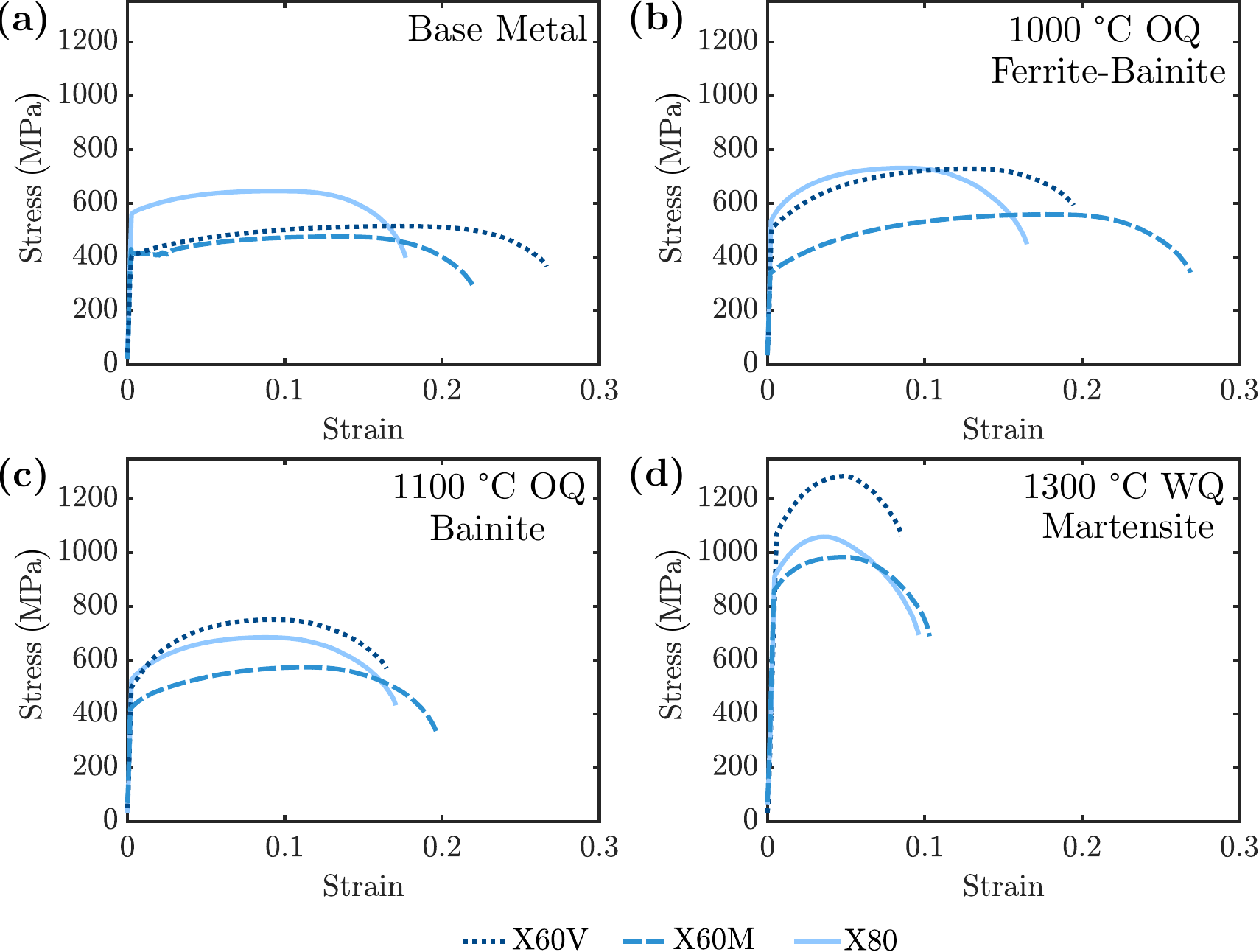}
\caption{Uniaxial stress-strain behaviour for each pipeline steel (X60V, X60M, X80) and microstructural condition considered: (a) base metal, (b) ferritic-bainitic microstructure (1000 $^\circ$C OQ), (c) bainitic microstructure (1100 $^\circ$C OQ), and (d) martensite (1300 $^\circ$C WQ).}
\label{fig:tensile_curves}
\end{figure}

The outputs from the tensile tests are given in Fig. \ref{fig:tensile_curves}, in the form of engineering stress-strain curves, with the yield strength $\text{R}_{p0.2}$, ultimate tensile strength $\sigma_{UTS}$, and strain at failure $\varepsilon_{\mathrm{f}}$ values being reported in Table \ref{tab:3}. All tests exhibited continuous yielding, except for the X60M base metal condition, which presented serrated yielding (see Fig. \ref{fig:tensile_curves}a). The base metal of both modern grades exhibited higher yield strength than the ferrite-bainite (1000 $^\circ$C OQ) samples. However, the ultimate tensile strength and ductility (measured as the strain at failure) were higher for the ferritic-pearlitic microstructures. Conversely, the vintage X60 steel presented lower yield strength and higher strain at failure in the base metal condition. The coarse-grain bainitic samples (1100 $^\circ$C OQ) of the vintage X60 steel showed higher yield strength when compared to the base metal and fine-grain ferritic-bainitic microstructures obtained for the same steel. After the heat treatment to produce a coarse-grain bainitic microstructure, the X80 steel exhibited yield strength and ductility values comparable to those of the X60V steel. The highest ductility was obtained for the modern X60 steel in the ferritic-bainitic condition (1000 $^\circ$C OQ). A notable increase in strength, and an associated drop in ductility, were observed for the martensite (1300 $^\circ$C WQ) condition. The highest yield strength was attained in the vintage X60 steel for the martensite condition, reaching a value of 1070 MPa. The mechanical tests also enabled estimating the yield strength and the ultimate tensile strength, as required for the fracture mechanics analysis (see Supplementary Material).\\

\subsection{Fracture toughness tests}

\subsubsection{Crack growth resistance in air}
\label{Sec:Kair}

\begin{figure}[H]
    \centering
    \includegraphics[width=\textwidth]{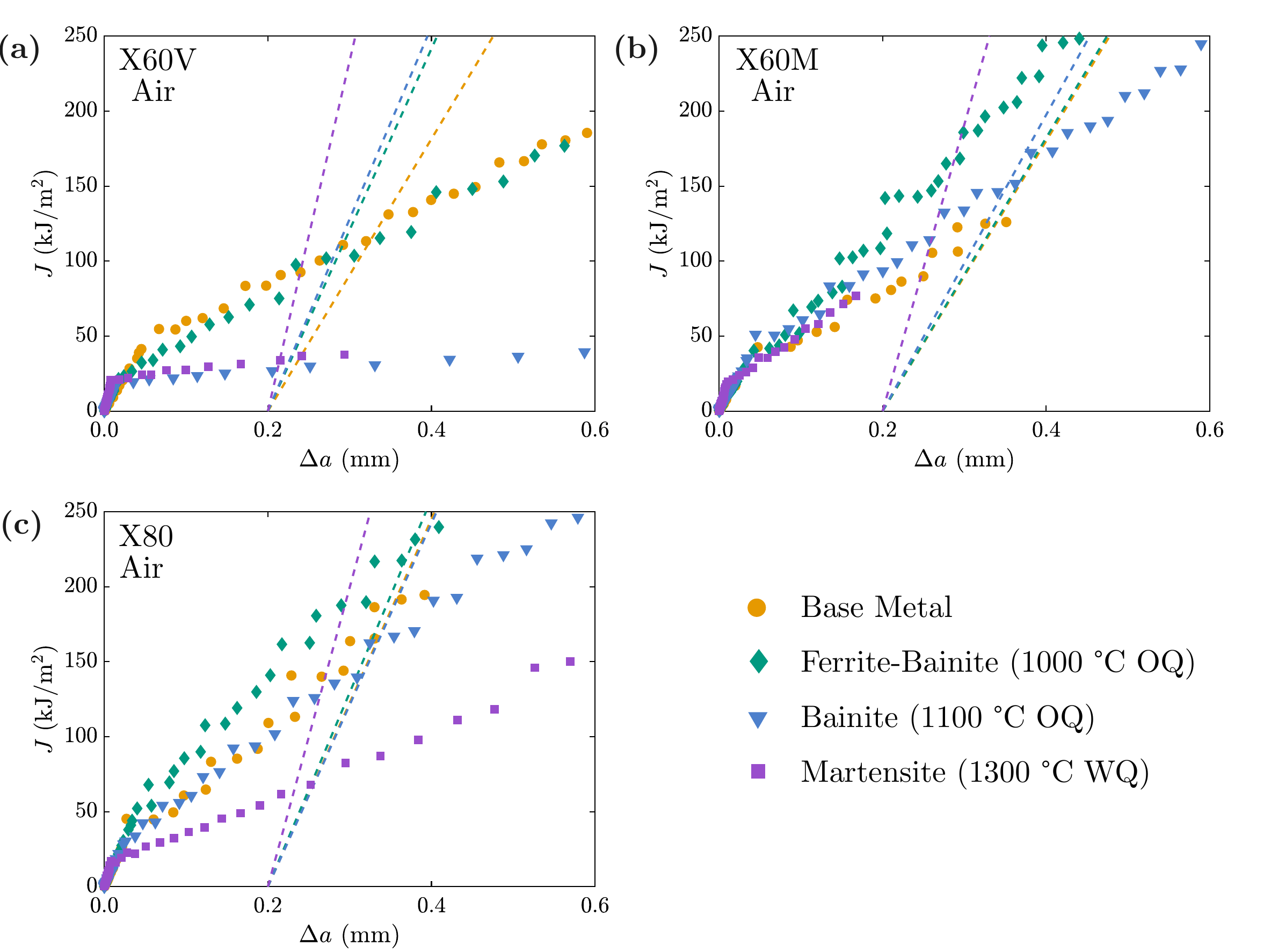}
    \caption{Crack growth resistance ($J$-$R$) curves obtained in air for the base metal, fine-grain microstructure (1000 $^\circ$C OQ), coarse-grain microstructure (1100 $^\circ$C OQ), and martensite-containing (1300 $^\circ$C WQ) microstructures for the three pipeline steels considered: (a) X60V, (b) X60M, and (c) X80. The dashed lines correspond to the 0.2 mm offset from the construction lines, to determine $J_{0.2}$.}
    \label{fig:JR_curves_Air}
\end{figure}

The outcome of the fracture tests in air is provided in Fig. \ref{fig:JR_curves_Air}, for the three steels (X60V, X60M, X80) and microstructure conditions considered. The base metal and the ferritic-bainitic microstructure cases exhibited similar behaviour and little sensitivity to the steel grade and age, with only a small reduction in crack growth resistance for the X60V case. However, the bainitic and martensitic microstructures exhibited a notable drop in crack growth resistance in the X60 vintage steel (Fig. \ref{fig:JR_curves_Air}a), with early unstable crack growth observed after an initial period of stable crack growth. A seemingly microstructurally-independent behaviour was attained for the modern grades (Fig. \ref{fig:JR_curves_Air}b), with some degradation in fracture performance observed for the martensite (1300 $^\circ$C WQ) condition in the X80 steel (Fig. \ref{fig:JR_curves_Air}c). Overall, the tearing resistance (slope of the $J$-$R$ curve) of modern steels was consistently greater than that of the vintage steel studied. An exception to this behavior was the base metal condition for the X60M steel case, possibly due to the effect of a higher fraction of pearlite as well as a high density of non-metallic MnS inclusions in the X60V steel that could have acted as ``crack arresters'' during crack propagation, as also suggested by Depraetere \textit{et al.} \cite{Depraetere2024TheSteel}.


\begin{figure}[H]
    \centering
    \includegraphics[width=\textwidth]{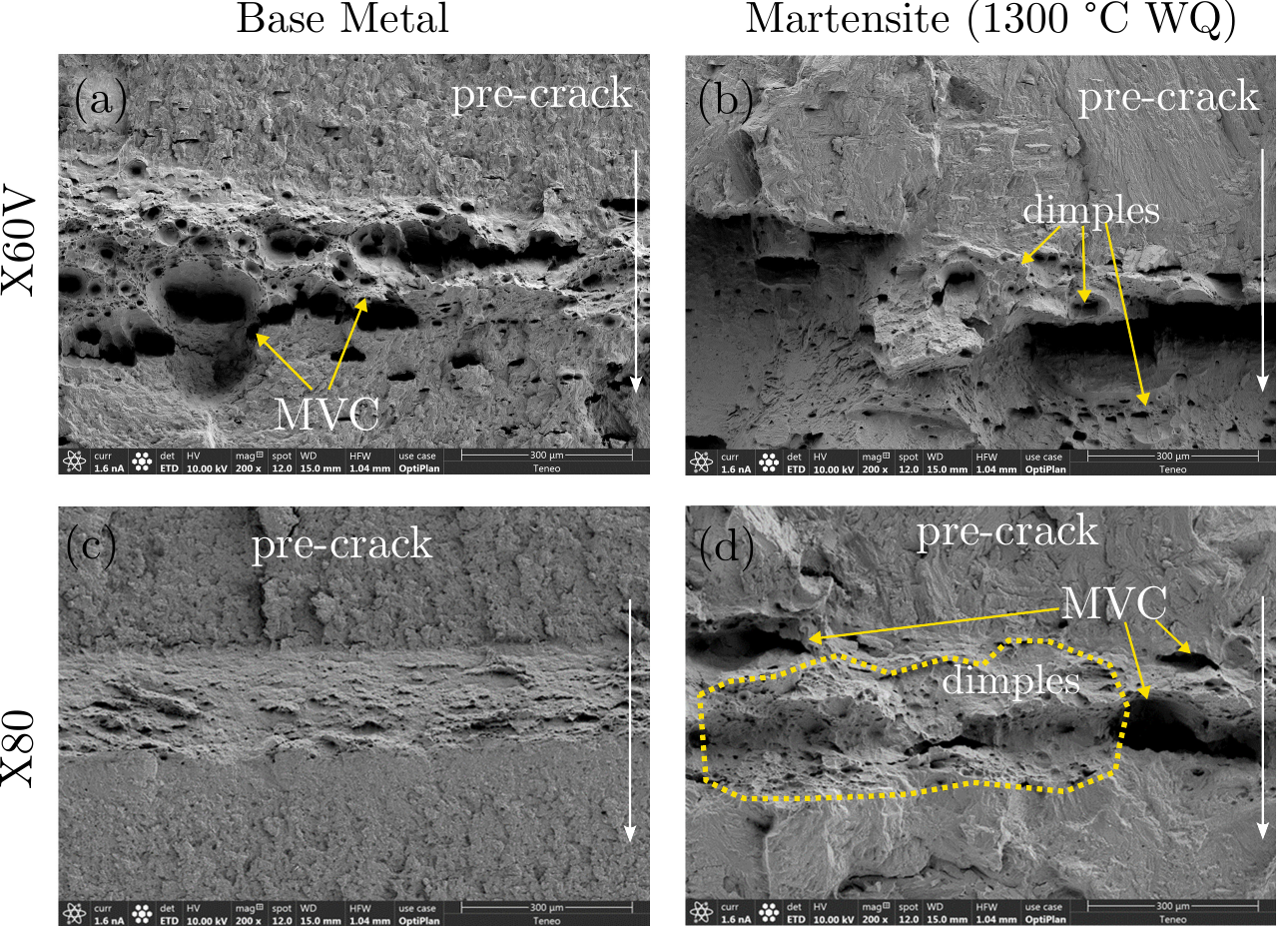}
    \caption{SEM micrographs of the fracture surfaces obtained after testing in air for the X60 vintage steel (top row) and the X80 modern steel (bottom row). The two limit cases are considered in terms of underlying microstructure: the base metal case (left column) and the martensite (1300 $^\circ$C WQ heat treatment) case (right column). The white arrow indicates the crack propagation direction and MVC stands for micro-void coalescence.}
    \label{fig:Air_FS}
\end{figure}

A selection of SEM micrographs of the fracture surfaces is provided to gain insight into the underlying mechanisms. First, consider the case of the vintage X60 steel, whose fracture surfaces are shown in Fig. \ref{fig:Air_FS}a for the base metal condition and in Fig. \ref{fig:Air_FS}b for the martensite (1300 $^\circ$C WQ) condition. Dimples are observed in both cases, with the base metal also showing micro-void coalescence (MVC) features, indicative of ductile fracture. This is in sharp contrast to the results obtained in H$_2$, which will be discussed next. Nevertheless, the significantly lower crack growth resistance attained for the X60V 1300 $^\circ$C WQ condition (see Fig. \ref{fig:JR_curves_Air}a) would suggest that the observed dimples are not necessarily an indication of a conventional ductile fracture process, but could instead be rationalised by the presence of a higher density of long strings of non-metallic MnS inclusions that act as crack initiation sites. In a soft matrix, such as the base metal case, these MnS inclusions would act instead as crack arresters \cite{Depraetere2024TheSteel,Chatzidouros2019FractureConditions}.


The fracture surfaces for the base metal and martensite (1300 $^\circ$C WQ) conditions in the case of the modern X80 steel are presented in Figs. \ref{fig:Air_FS}c and \ref{fig:Air_FS}d, respectively. Both dimples and microvoid coalescence (MVC) occurrences are observed for both microstructural conditions. The dimples observed on the base metal fracture surface in X80 are smaller in size and density, compared to X60V. These observations are consistent with the higher fracture resistance reported in the $J$-$R$ curves for the X80 case, relative to the X60V one (across all microstructures, see Fig. \ref{fig:JR_curves_Air}). 


\subsubsection{Crack growth resistance in H$_2$}

\begin{figure}[H]
    \centering
    \includegraphics[width=\textwidth]{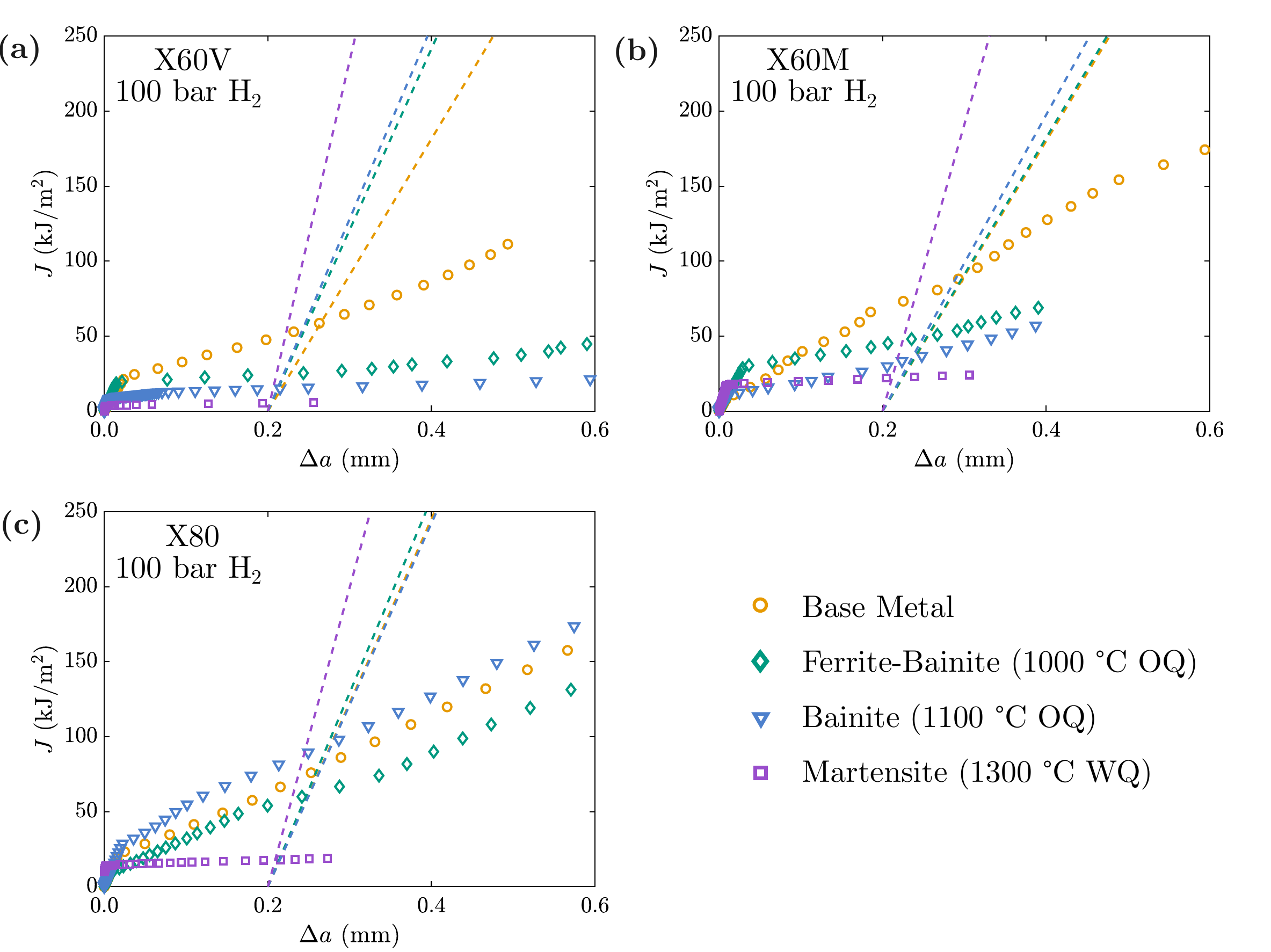}
    \caption{Crack growth resistance ($J$-$R$) curves obtained in 100 bar H$_\text{2}$ gas for the base metal, ferritic-bainitic microstructure (1000 $^\circ$C OQ), bainitic microstructure (1100 $^\circ$C OQ), and martensite-containing (1300 $^\circ$C WQ) microstructures for the three pipeline steels considered: (a) X60V, (b) X60M, and (c) X80. The dashed lines correspond to the 0.2 offset from the construction lines, to determine $J_{0.2}$.}
    \label{fig:JR_curves_H2}
\end{figure}

When exposed to 100 bar of pure H$_2$, all the steels and microstructures investigated exhibited a significant reduction in fracture resistance, as exemplified in the $J$-$R$ curves provided in Fig. \ref{fig:JR_curves_H2}. Firstly, an influence of microstructure was observed on the hydrogen-assisted cracking resistance for each steel. On the one hand, the base metal condition revealed the highest crack growth resistance for both the vintage and modern X60 grade. On the other hand, the simulated HAZ microstructures exhibited a progressive decrease in fracture resistance with increasing grain size. In other words, coarser microstructures exhibited the lowest resistance to fracture in hydrogen, along with a pronounced decrease in the slope of the R-curve compared to tests conducted in air. Secondly, a notable effect of the steelmaking era was observed in the tearing modulus (d$J$/d$a$) of the R-curves obtained in hydrogen. For example, the modern X60 showed higher crack growth resistance than the vintage X60 steel for the base metal, ferritic-bainitic, and coarse bainitic microstructures. In addition, unstable crack propagation was observed for the coarse bainitic condition of the vintage X60 steel. All the martensite-containing specimens presented unstable cracking in hydrogen, irrespective of steel age or grade. The least fracture-resistant behaviour observed among all the conditions considered is that of a martensite microstructure in a vintage X60 pipeline, the case considered here as representative of a plausible worst-case scenario.

\begin{figure}[H]
    \centering
    \includegraphics[width=\textwidth]{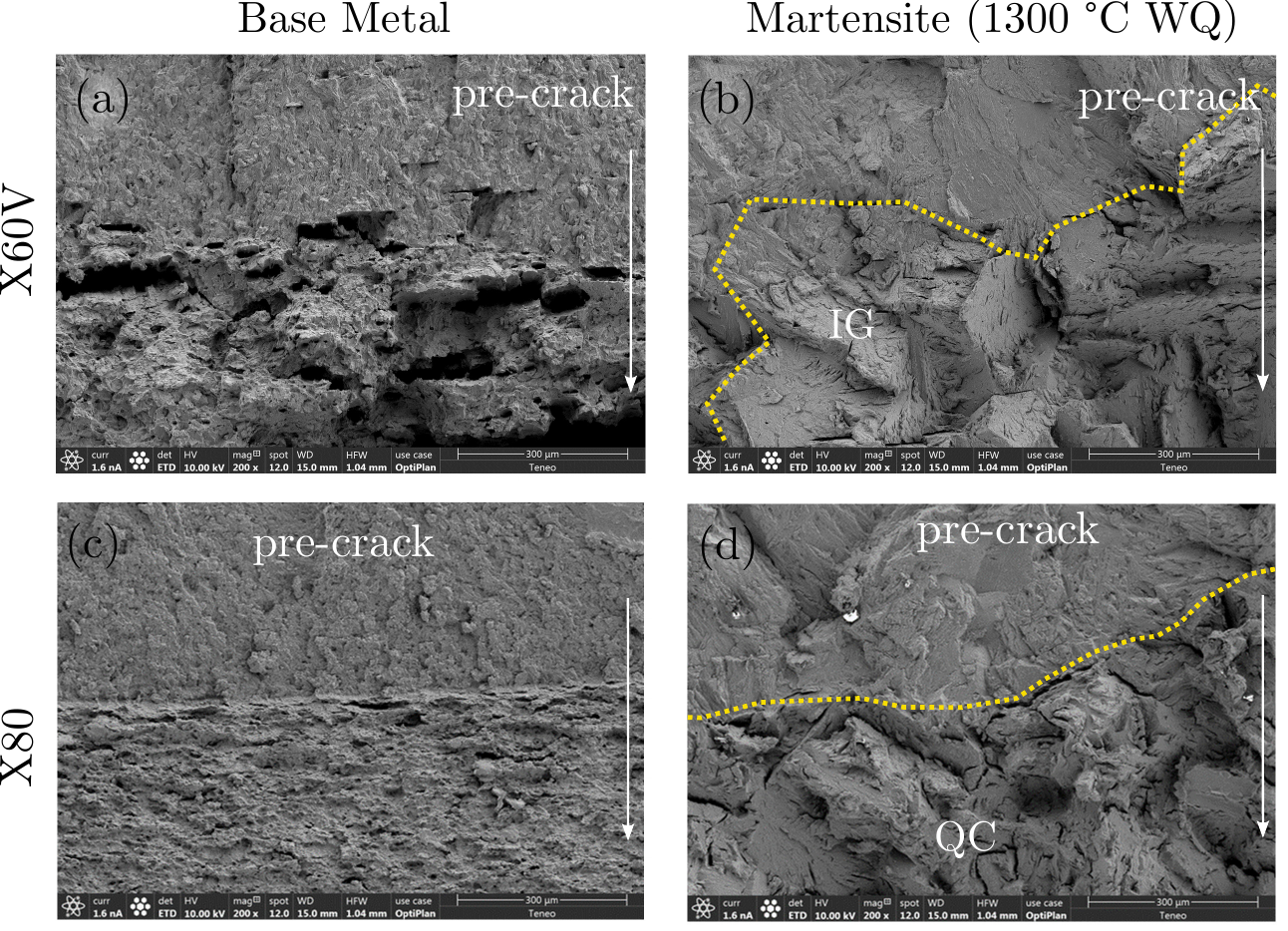}
    \caption{SEM micrographs of the fracture surfaces obtained after testing in 100 bar H$_\text{2}$ gas for the X60 vintage steel (top row) and the X80 modern steel (bottom row). The two limit cases are considered in terms of underlying microstructure: the base metal case (left column) and the martensite (1300 $^\circ$C WQ heat treatment) case (right column). The white arrow indicates the crack propagation direction, IG stands for intergranular cracking, and QC fo quasi-cleavage.}
    \label{fig:H2_FS}
\end{figure}

The $J$-$R$ findings are supported by SEM micrographs of the fracture surfaces, as shown in Fig. \ref{fig:H2_FS}. The fracture surface of the base metal of X60V tested in hydrogen exhibited a stepped morphology, indicating crack propagation through successive planes, as observed in Fig. \ref{fig:H2_FS}a. In the fully martensitic condition, X60V showed a morphology consistent with intergranular cracking (IG), characteristic of the hydrogen-assisted fracture process (Fig. \ref{fig:H2_FS}b). In contrast, the base metal of the X80 steel presented a similar fracture surface in air and hydrogen (see Figs. \ref{fig:Air_FS}c and \ref{fig:H2_FS}c), while the martensite-containing material exhibited quasi-cleavage (QC) characteristics after exposure to hydrogen (see Fig. \ref{fig:H2_FS}d).

\begin{figure}[H]
    \centering
    \includegraphics[width=\textwidth]{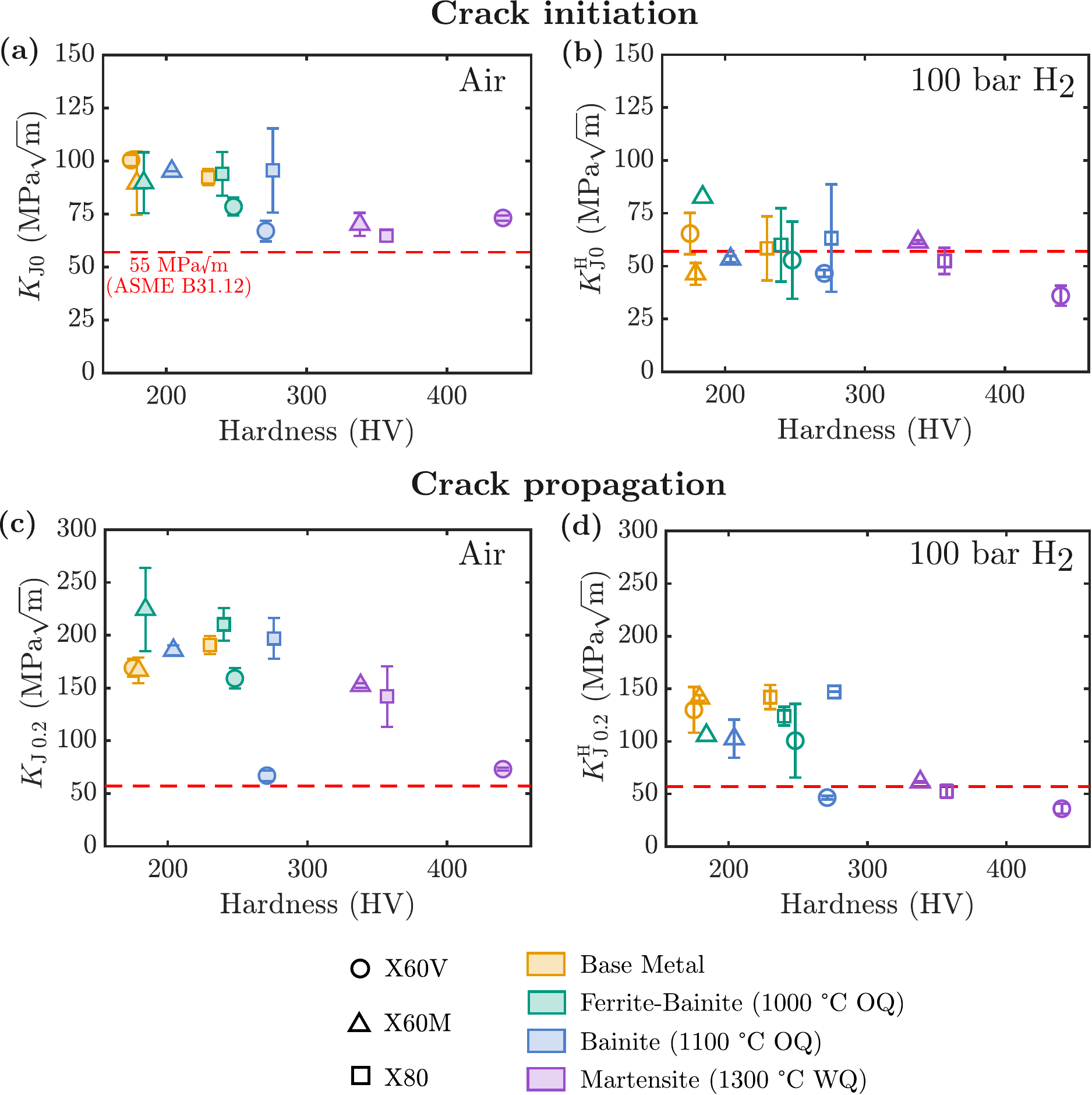}
    \caption{Fracture toughness calculated for crack initiation, K$_\text{J0}$ (top) and crack propagation, K$_\text{J0.2}$ (bottom) versus the material hardness in air, (a) and (c) (left column), and in 100 bar H$_\text{2}$ gas, (b) and (d) (right column). The red dashed line corresponds to the fracture toughness limit for hydrogen pipelines specified in the ASME B31.12 standard \cite{Asme2023HydrogenB31.12-2023}.}
    \label{fig:K_vs_Hardness}
\end{figure}

A summary of the results obtained is presented in Fig. \ref{fig:K_vs_Hardness}, where the critical stress intensity factor at initiation $K_{J0}$ and growth $K_{J0.2}$ are presented for all the conditions considered (three steels and four microstructures), in air and 100 bar H$_2$ environments. The results are shown as a function of hardness, so as to better illustrate the influence of the microstructure and the interplay between hardness and hydrogen embrittlement susceptibility. In all cases, a notable drop in both $K_{J0}$ and $K_{J0.2}$ is observed when the samples are exposed to hydrogen, in agreement with expectations and literature findings for base material behaviour. Importantly, while the base material of the vintage steel exhibited an initiation toughness above the current ASME B31.12 limit (55 MPa$\sqrt{\text{m}}$), presented in the standard as indicative of safe operation conditions, all the HAZ microstructures resulted in values of $K_{J0}^H$ below this acceptance limit. The performance was particularly poor for the martensitic microstructure in the vintage steel, resulting in an initiation toughness of $K_{J0}^H=32$ MPa$\sqrt{\text{m}}$, which emphasises the higher susceptibility of hard, brittle zones within the HAZ. The results obtained for the 0.2 mm fracture toughness (Figs. \ref{fig:K_vs_Hardness}c and d) are qualitatively similar to those of the resistance to crack initiation, but with higher toughness values overall. Nevertheless, once again, scenarios exist that result in a toughness value below the ASME B31.12 acceptance criterion of 55 MPa$\sqrt{\text{m}}$; this is the case for the coarse-grain and martensite-dominated microstructures for the X60V and of the martensite-dominated microstructure for the X80. In any case, care should be taken when determining safe operation across a wide range of conditions using a single toughness value; the use of Failure Assessment Diagrams (FAD) would provide a more precise assessment \cite{wijnen2025}.

\begin{figure}[H]
    \centering
    \includegraphics[width=\textwidth]{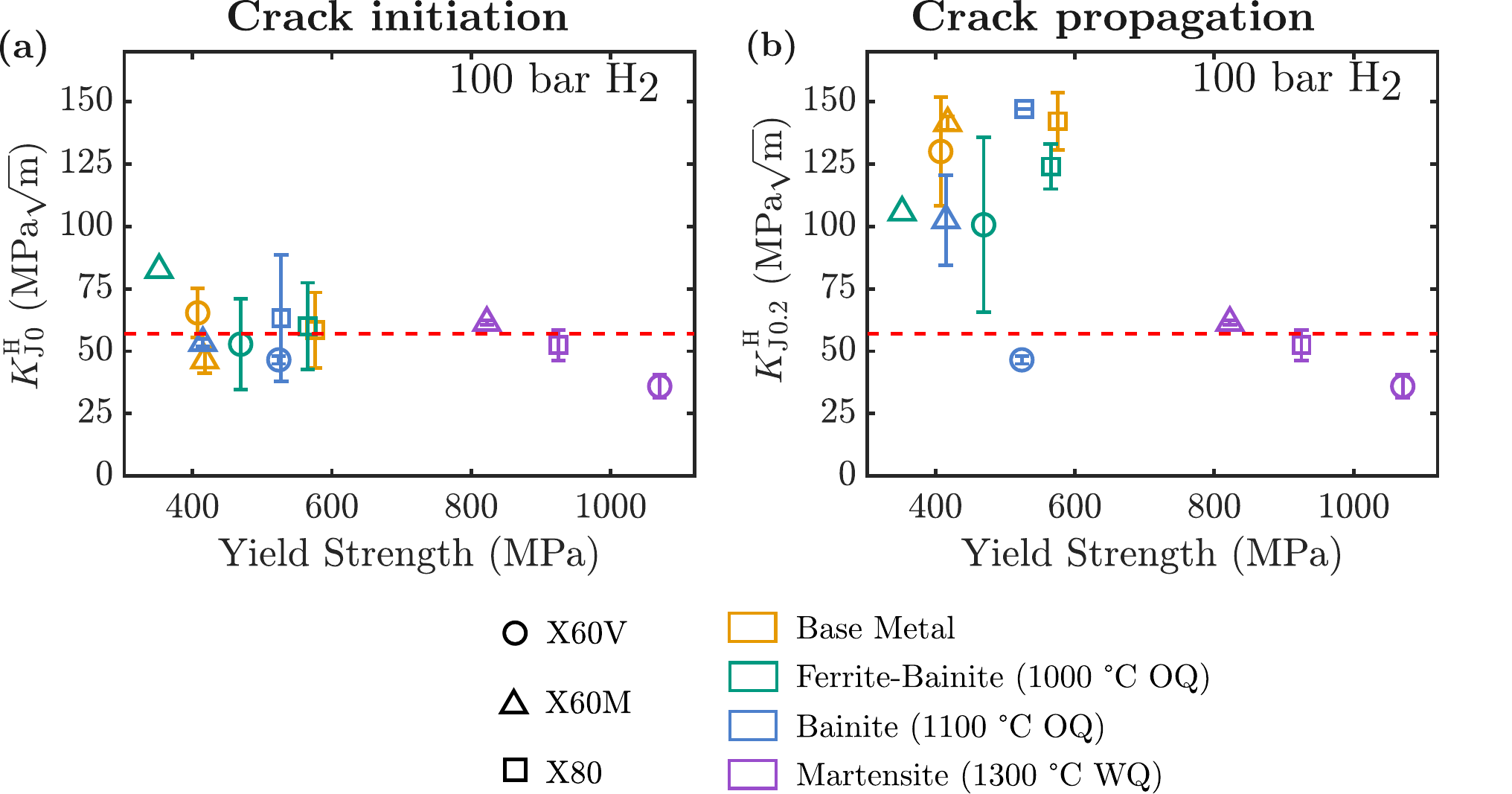}
    \caption{Fracture toughness calculated for crack initiation, K$_\text{J0}$ (a) and crack propagation, K$_\text{J0.2}$ (b) versus the yield strength in 100 bar H$_\text{2}$ gas. The red dashed line corresponds to the fracture toughness limit for hydrogen pipelines specified in the ASME B31.12 standard \cite{Asme2023HydrogenB31.12-2023}.}
    \label{fig:FT_vs_YS}
\end{figure}

Some degree of correlation between hardness and hydrogen embrittlement susceptibility is observed, particularly for the crack growth resistance measurements. Nevertheless, the results reveal multiple conditions where hardness alone is not sufficient to assess the susceptibility to hydrogen embrittlement, emphasising the need for microstructural characterisation. This is, for example, observed in the case of the coarse bainite-containing microstructures (1100 $^\circ$C OQ treatment) in the X60V and X80 steels - while they result in similar hardness levels ($\sim$ 270 HV), the vintage samples resulted in unstable cracking and significantly lower toughness values (three times smaller in the case of $K_{J0.2}^H$). A similar trend is observed when the results are plotted as a function of the yield strength, as shown in Fig. \ref{fig:FT_vs_YS} for the H$_2$ tests. While the lowest resistance to fracture corresponds to the martensitic microstructures, which are associated with very high yield strength values, early cracking (below the ASME acceptance limit) is also observed for mid-strength cases. 

\begin{figure}[H]
    \centering
    \includegraphics[width=0.9\textwidth]{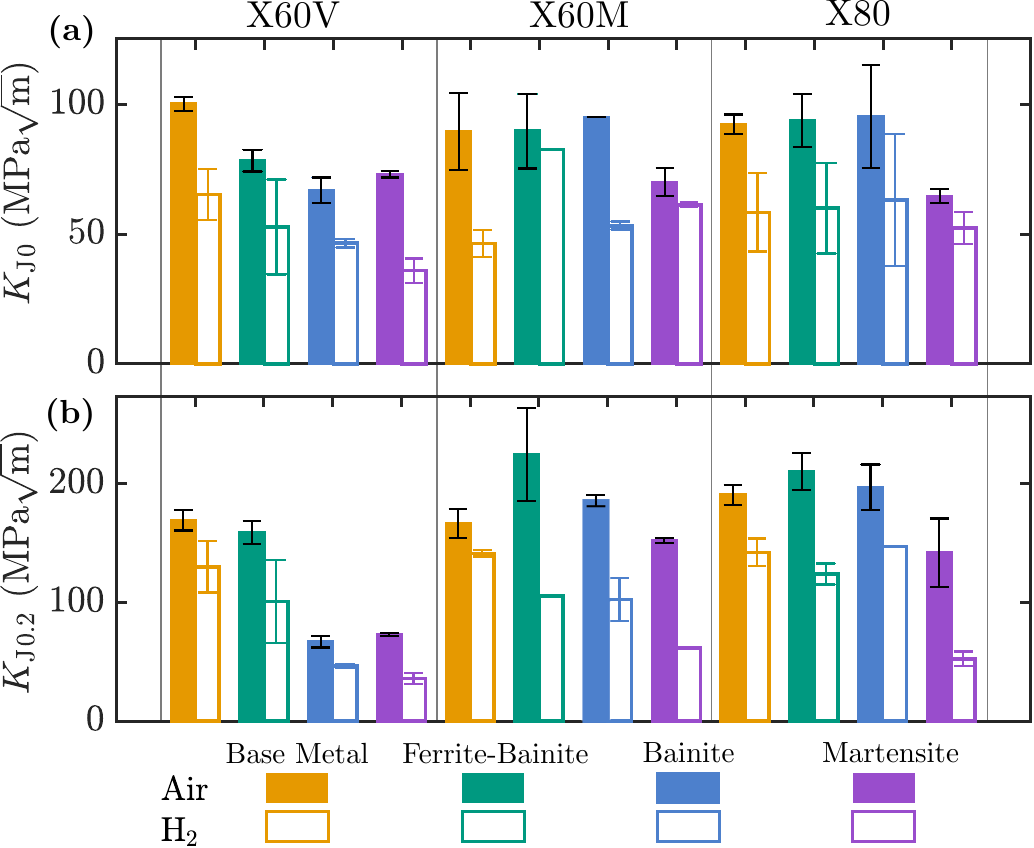}
    \caption{Summary of the crack initiation (a) and growth (b) toughnesses measured in air (filled bars) and in 100 bar H$_\text{2}$ gas (unfilled bars) for all the conditions considered: three pipeline steels (X60V - left column, X60M - centre column, X80 - right column) and four microstructural conditions (base metal - orange, ferrite-bainite - green, bainite - blue, martensite - purple).}
    \label{fig:bar_chart}
\end{figure}

An overview of the results obtained is given in the bar chart presented in Fig. \ref{fig:bar_chart}. Both crack initiation and growth toughness measurements are provided, for air and H$_2$ environments. It must be highlighted that the $K_{J0.2}$ measurements for base metal are in good agreement with the literature. For example, San Marchi \textit{et al.} \cite{SanMarchi2010FractureHydrogen} reported $K_{J0.2}$ values between 77 and 131 MPa$\sqrt{\text{m}}$ for modern X60 and X70/X80 steels in gaseous hydrogen between 55 bar and 210 bar.  

In the presence of hydrogen, the initiation resistance showed greater sensitivity to the microstructural condition than to the steel era. For example, the base metal X60V case reveals a higher initiation resistance than any other microstructure in both the X60V and X60M steels.
In general, the vintage material exhibited a lower fracture resistance than the modern steels. This is particularly evident when looking at the $K_{J0.2}$ measurements for the various simulated weld microstructures and the hydrogen environment. The higher crack propagation resistance observed in modern steels may be attributed to their improved cleanliness, particularly due to their lower content of non-metallic inclusions. Among these, elongated MnS are of particular concern, as they can facilitate cracking in hydrogen-containing environments \cite{Chatzidouros2019FractureConditions}. Elongated MnS inclusions with lengths exceeding 50 \si{\micro\metre} were observed in the vintage steel examined in this study (Fig.~\ref{fig:bm_microstructures} (a)). Furthermore, the results show that the steel manufacturing era can play a notable role by promoting microstructures more susceptible to hydrogen embrittlement, mainly due to the higher carbon content and increased grain coarsening observed during thermal cycles. This behaviour is illustrated by the martensitic-containing microstructure of X60V, which showed comparable crack initiation resistance in air to that of the modern steels, but exhibited the lowest fracture toughness under hydrogen among all conditions studied.\\

\begin{figure}[H]
    \centering
    \includegraphics[width=0.76\textwidth]{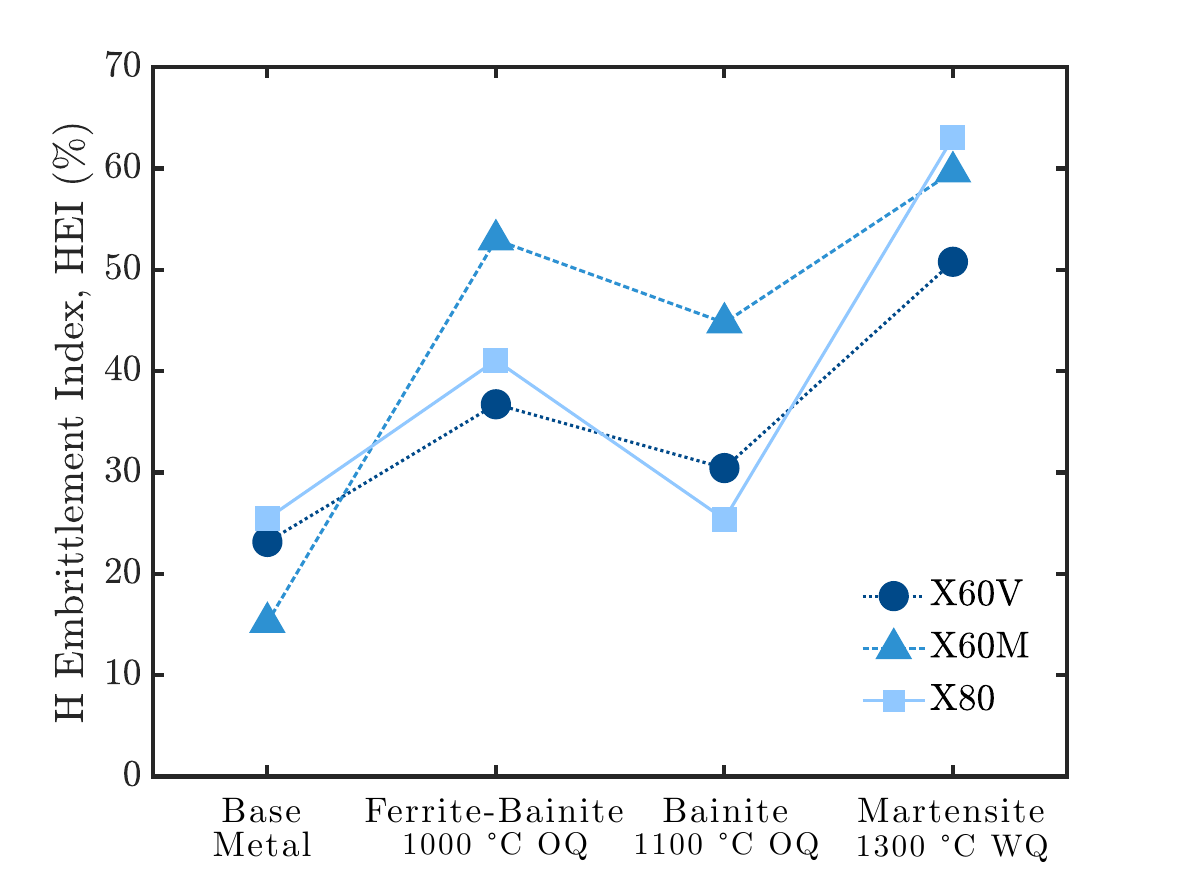}
    \caption{Hydrogen embrittlement index (HEI), calculated from the fracture toughness ($K_{J0.2}$) measurements in 100 bar H$_\text{2}$ gas using Eq. (\ref{eq:HEI}). The results are shown as a function of the microstructural conditions and steels considered.}
    \label{fig:EI}
\end{figure}

The interplay between microstructure, steel characteristics (age, grade, composition), and hydrogen embrittlement susceptibility can be more easily visualised by defining a hydrogen embrittlement index (HEI), from the fracture toughness values measured in air ($K_{J0.2}$) and in 100 bar H$_2$ gas ($K_{J0.2}^H$):
\begin{equation}
\text{HEI} = \frac{K_{J0.2}-K_{J0.2}^H}{K_{J0.2}} \cdot 100
\label{eq:HEI}
\end{equation}
\noindent such that 0\% represents no change in the fracture toughness (i.e., no embrittlement) and 100\% indicates that the material has reached a full toughness reduction due to hydrogen embrittlement. The variation of this embrittlement index with microstructure and pipeline steel is shown in Fig. \ref{fig:EI}. 

All the steels show the same trend, regardless of their manufacturing era. The base metal exhibits the lowest degree of embrittlement susceptibility, with the martensitic condition displaying the highest vulnerability, and the fine- and coarse-grain microstructures being at an intermediate level. The significant influence that the microstructure can have becomes evident, with the embrittlement index going from $\sim$20\% for the base metal to approximately 50-60\% for the martensite condition. Comparing across steel grades and age, it appears that the X60M exhibited the largest susceptibility to HAZ microstructure and hydrogen, despite outperforming X60V in terms of overall resistance to hydrogen-assisted cracking, as observed in Fig.~\ref{fig:bar_chart}. This highlights that, while the lowest toughness values are attained for the vintage steel, modern steel grades can exhibit a significant susceptibility to microstructural heterogeneities and hydrogen (HEI $>$25\% in all cases, and larger than 60\% for the martensitic case). Nevertheless, the relative information provided by the HEI graph cannot be used to draw design implications. For example, while the X60M ferritic-bainitic (1000 $^\circ$C OQ) condition resulted in a HEI larger than 50\%, the fracture toughness in 100 bar H$_2$ is approximately 100 MPa$\sqrt{\text{m}}$, which is well above the minimum threshold value stipulated in ASME B31.12. Thus, because HEI only represents a relative value between in-air and in-hydrogen properties with no consideration given to the starting value, it should not be used as a decision-making tool for establishing the suitability of a pipeline steel for hydrogen service.\\

The quantitative toughness information gained in this study can be directly used for design and fitness-for-service assessment when a single microstructure governs pipeline integrity, with the martensite case being a conservative choice. However, in many scenarios, weld integrity will be governed by the composite behaviour of multiple microstructures. In such cases, the use of computational models that can predict macroscopic behaviour while resolving the weld microstructures is needed \cite{wijnen2025computational}. Such models require individual microstructural information and, therefore, are enabled by the protocol and results obtained in this work. Future work will showcase how the combination of individual microstructural insight and microstructurally-informed models can be used to predict pipeline integrity for arbitrary choices of service conditions, material and weld protocol.

\section{Conclusions}
\label{sec6}

We have presented a new protocol to quantitatively characterise the resistance to hydrogen-assisted fracture of the different microstructural regions present within the heat-affected zone (HAZ) of natural gas pipelines. This is of pressing importance, given the short-term plan of repurposing existing natural gas pipelines to transport hydrogen and the higher vulnerability of these regions, which will govern the structural integrity of hydrogen pipeline systems. Tailored heat treatments are conducted on pipeline steels spanning different manufacturing ages (vintage vs modern), grades, and compositions to replicate the microstructures presented in the various sub-zones of the HAZ. Specifically, base metal performance is benchmarked against a fine-grain ferritic-bainitic microstructure, a coarse-grain bainite microstructure, and a mainly martensitic microstructure. The three microstructures considered are relevant to those observed in welds of pipelines taken out of service. The application of these tailored heat treatments enabled obtaining homogeneous, bulk samples that could be used for quantitative fracture mechanics testing. Fracture tests were conducted in air and in a 100-bar pure H$_2$ environment for all the combinations of pipeline steel (X60 vintage, X60 modern, X80) and microstructure (base metal, ferrite-bainite, bainite, martensite). The main findings include:
\begin{itemize}
    \item Tailored heat treatments can successfully simulate weld HAZ regions, providing testing-scale samples with relevant microstructures, mechanical behaviour, and hardness levels.

    \item In all cases, the exposure to 100 bar of pure H$_2$ resulted in a notable drop in fracture resistance and a shift in the morphology of the fracture surface, from ductile-like features to brittle ones.

    \item All the simulated HAZ microstructures exhibited a higher degree of hydrogen embrittlement susceptibility than the base metal, emphasising the importance of their independent characterisation.

    \item An effect of the manufacturing era was observed, with vintage steels exhibiting lower fracture resistance in both in-air and H$_2$ environments, due to higher non-metallic inclusion content and carbon content.

    \item The fracture toughness measured in 100 bar H$_2$ was found to be below the current ASME B31.12 acceptability threshold (55 MPa$\sqrt{\text{m}}$) for various microstructures (coarse-grain bainite, martensite) and steels (X60 vintage, X80). This is in sharp contrast with the base metal behaviour, which rendered fracture toughness values at least twice as high.\

    \item The microstructural influence was found to dominate over steel age, composition, and grade in terms of initiation to fracture resistance. 

    \item While some degree of correlation between resistance to hydrogen-assisted fracture and hardness can be inferred, differences in toughness measurements are observed for steel/microstructure combinations that result in similar hardness, emphasising the need for the consideration of additional criteria.

    \item The lowest fracture initiation resistance (32 MPa$\sqrt{\text{m}}$) was attained for the combination of the X60 vintage steel and the martensitic microstructure, a condition representative of a plausible worst-case scenario and thus what can potentially govern pipeline failure.
\end{itemize}

These findings highlight the need for an integral structural integrity assessment of hydrogen transport pipelines, including not only base material behaviour and hardness, but also microstructure and steel characteristics. 

\section*{Acknowledgments}
\label{Acknowledge of funding}

\noindent The authors acknowledge the support and assistance of Dr A. D\'{\i}az and Dr I.I. Cuesta (University of Burgos) with the H$_2$ testing. D. Chalfoun acknowledges the assistance of Dr J. Wijnen (University of Oxford) in the determination of the heat treatment cooling rates through numerical modelling. The authors acknowledge financial support from EPRI through the R\&D project ``Virtual Testing of hydrogen-sensitive components''. E.\ Mart\'{\i}nez-Pa\~neda acknowledges financial support from UKRI's Future Leaders Fellowship programme [grant MR/V024124/1] and from the UKRI Horizon Europe Guarantee programme (ERC Starting Grant \textit{ResistHfracture}, EP/Y037219/1).



\small
\begin{thebibliography}{10}
\expandafter\ifx\csname url\endcsname\relax
  \def\url#1{\texttt{#1}}\fi
\expandafter\ifx\csname urlprefix\endcsname\relax\def\urlprefix{URL }\fi
\expandafter\ifx\csname href\endcsname\relax
  \def\href#1#2{#2} \def\path#1{#1}\fi

\bibitem{telessy2024repurposing}
K.~T{\'e}lessy, L.~Barner, F.~Holz, {Repurposing natural gas pipelines for
  hydrogen: Limits and options from a case study in Germany}, International
  Journal of Hydrogen Energy 80 (2024) 821--831.

\bibitem{sayani2025techno}
J.~K.~S. Sayani, M.~Wang, Z.~Ma, P.~Sharan, M.~Mehana, B.~Chen, Techno-economic
  analysis of hydrogen transport via repurposed natural gas pipelines: Flow
  dynamics and infrastructure tradeoffs, International Journal of Hydrogen
  Energy 136 (2025) 789--821.

\bibitem{guzzo2025hydrogen}
G.~Guzzo, I.~Saedi, S.~Mhanna, C.~Carcasci, P.~Mancarella, Hydrogen blending in
  gas pipelines: Fluid-dynamic insights, risks, and recommendations,
  International Journal of Hydrogen Energy 120 (2025) 67--77.

\bibitem{djukic2019synergistic}
M.~B. Djukic, G.~M. Bakic, V.~S. Zeravcic, A.~Sedmak, B.~Rajicic, The
  synergistic action and interplay of hydrogen embrittlement mechanisms in
  steels and iron: Localized plasticity and decohesion, Engineering Fracture
  Mechanics 216 (2019) 106528.

\bibitem{malheiros2022local}
L.~C. Malheiros, A.~Oudriss, S.~Cohendoz, J.~Bouhattate, F.~Th{\'e}bault,
  M.~Piette, X.~Feaugas, Local fracture criterion for quasi-cleavage
  hydrogen-assisted cracking of tempered martensitic steels, Materials Science
  and Engineering: A 847 (2022) 143213.

\bibitem{yu2024hydrogen}
H.~Yu, A.~D{\'\i}az, X.~Lu, B.~Sun, Y.~Ding, M.~Koyama, J.~He, X.~Zhou,
  A.~Oudriss, X.~Feaugas, et~al., {Hydrogen embrittlement as a conspicuous
  material challenge--Comprehensive review and future directions}, Chemical
  Reviews 124~(10) (2024) 6271--6392.

\bibitem{chen2024hydrogen}
Y.-S. Chen, C.~Huang, P.-Y. Liu, H.-W. Yen, R.~Niu, P.~Burr, K.~L. Moore,
  E.~Mart{\'\i}nez-Pa{\~n}eda, A.~Atrens, J.~M. Cairney, {Hydrogen trapping and
  embrittlement in metals-- A review}, International Journal of Hydrogen Energy
  147 (2025) 150033.

\bibitem{ronevich2018fatigue}
J.~A. Ronevich, C.~R. D'Elia, M.~R. Hill, {Fatigue crack growth rates of X100
  steel welds in high pressure hydrogen gas considering residual stress
  effects}, Engineering Fracture Mechanics 194 (2018) 42--51.

\bibitem{jemblie2024safe}
L.~Jemblie, A.~B. Hagen, C.~H.~M. Hagen, B.~Nyhus, A.~Alvaro, D.~Wang, E.~A.
  Koren, R.~Johnsen, Z.~Zhang, J.~Yamabe, et~al., Safe pipelines for hydrogen
  transport, International Journal of Hydrogen Energy 136 (2025) 672--685.

\bibitem{Depraetere2024TheSteel}
R.~Depraetere, W.~De~Waele, M.~Cauwels, T.~Depover, K.~Verbeken,
  S.~Hertel{\'{e}}, {The effect of hydrogen and notch orientation in SENT
  specimens on the fracture toughness of an API 5L X70 pipeline steel},
  Engineering Fracture Mechanics 300 (4 2024).

\bibitem{Ronevich2021Hydrogen-assistedHydrogen}
J.~A. Ronevich, E.~J. Song, B.~P. Somerday, C.~W. San~Marchi,
  {Hydrogen-assisted fracture resistance of pipeline welds in gaseous
  hydrogen}, International Journal of Hydrogen Energy 46 (2021) 7601--7614.

\bibitem{tahan2022recent}
M.-R. Tahan, Recent advances in hydrogen compressors for use in large-scale
  renewable energy integration, International Journal of Hydrogen Energy
  47~(83) (2022) 35275--35292.

\bibitem{mandal2024computational}
T.~K. Mandal, J.~Parker, M.~Gagliano, E.~Mart{\'\i}nez-Pa{\~n}eda,
  Computational predictions of weld structural integrity in hydrogen transport
  pipelines, International Journal of Hydrogen Energy 136 (2025) 923--937.

\bibitem{zhang2024modeling}
J.~Zhang, Y.~F. Cheng, Modeling of hydrogen atom distribution at corrosion
  defect on existing pipelines repurposed for hydrogen transport under pressure
  fluctuations, International Journal of Hydrogen Energy 58 (2024) 1075--1087.

\bibitem{brown2022development}
D.~Brown, K.~Reddi, A.~Elgowainy, The development of natural gas and hydrogen
  pipeline capital cost estimating equations, International Journal of Hydrogen
  Energy 47~(79) (2022) 33813--33826.

\bibitem{vreeburg2025potential}
J.~R. Vreeburg, J.~C. Garcia-Navarro, The potential of repurposing offshore
  natural gas infrastructure on the dutch continental shelf for hydrogen
  production and transport, International Journal of Hydrogen Energy 115 (2025)
  37--48.

\bibitem{mielich2025europe}
T.~Mielich, R.~Dobler, J.~Hollnagel, O.~Akca, J.~M{\"u}ller-Kirchenbauer,
  Europe's way from natural gas to green hydrogen: Modeling and simulation of
  the transforming european gas transport infrastructure, International Journal
  of Hydrogen Energy 135 (2025) 156--171.

\bibitem{Ronevich2024INFLUENCESTEELS}
J.~Ronevich, M.~Agnani, M.~Gagliano, J.~Parker, C.~S. Marchi, {Influence of
  hardness on hydrogen-assisted fracture in pipeline steels}, in: Proceedings
  of the 2024 15th International Pipeline Conference, Calgary, Alberta, Canada,
  2024, pp. IPC2024--133937.

\bibitem{Kappes2023HydrogenMethods}
M.~A. Kappes, T.~Perez, {Hydrogen blending in existing natural gas transmission
  pipelines: A review of hydrogen embrittlement, governing codes, and life
  prediction methods}, Corrosion Reviews 41 (2023) 319--347.

\bibitem{Leis2015ManagingInfrastructure}
B.~N. Leis, Managing an aging pipeline infrastructure, in: R.~W. Revie (Ed.),
  Oil and Gas Pipelines: Integrity and Safety Handbook, John Wiley \& Sons,
  Inc., 2015, Ch.~43, pp. 609--634.

\bibitem{API5L}
{American Petroleum Institute}, {Line Pipe: API Specification 5L} (4 2018).

\bibitem{Gangloff2003}
R.~P. Gangloff, Hydrogen-assisted cracking, in: I.~Milne, R.~Ritchie,
  B.~Karihaloo (Eds.), Comprehensive Structural Integrity Vol. 6, Elsevier
  Science, 2003, pp. 31--101.

\bibitem{nanninga2010role}
N.~Nanninga, J.~Grochowsi, L.~Heldt, K.~Rundman, Role of microstructure,
  composition and hardness in resisting hydrogen embrittlement of fastener
  grade steels, Corrosion Science 52~(4) (2010) 1237--1246.

\bibitem{wijnen2025computational}
J.~Wijnen, J.~Parker, M.~Gagliano, E.~Mart{\'\i}nez-Pa{\~n}eda, A computational
  framework to predict weld integrity and microstructural heterogeneity:
  Application to hydrogen transmission, Materials \& Design 249 (2025) 113533.

\bibitem{wijnen2025}
J.~Wijnen, J.~Parker, M.~Gagliano, E.~Mart{\'\i}nez-Pa{\~n}eda, Virtual failure
  assessment diagrams for hydrogen transmission pipelines, International
  Journal of Hydrogen Energy (in press) (2025).

\bibitem{Alvaro2014}
A.~Alvaro, V.~Olden, A.~Macadre, {Hydrogen embrittlement susceptibility of a
  weld simulated X70 heat affected zone under H2 pressure}, Materials Science
  and Engineering: A 597 (2014) 29--36.

\bibitem{Madi2024MechanicalSpecimens}
Y.~Madi, L.~M. Santana, S.~Belkacemi, V.~Farrugia, A.~Meddour, P.~J. Marchais,
  M.~Bertin, J.~Furtado, {Mechanical characterization of hydrogen embrittlement
  in a gaseous environment: An innovative test setup using sub-size specimens},
  Engineering Failure Analysis 162 (2024) 108362.

\bibitem{Bortot2024InvestigationEnvironment}
P.~Bortot, M.~Ortolani, M.~Sileo, E.~Escorza, M.~Connolly, Z.~N. Buck,
  A.~Chandra, {Investigation of the fracture resistance of high strength
  ferritic steel welds in gaseous hydrogen environment}, International Journal
  of Hydrogen Energy 136 (2025) 777--788.

\bibitem{ASTM2024ASTMSize}
{ASTM}, {ASTM E12-24: Test Methods for Determining Average Grain Size} (2024).

\bibitem{James2015FailurePipelines}
B.~James, A.~Hudgins, {Failure analysis of oil and gas transmission pipelines},
  in: Handbook of Materials Failure Analysis with Case Studies from the Oil and
  Gas Industry, Elsevier, 2015, pp. 1--38.

\bibitem{Stearling1992IntroductionWelding}
K.~Stearling, Introduction to the Physical Metallurgy of Welding, 2nd Edition,
  Elsevier, Oxford, 1992.

\bibitem{Zerbst2014ReviewPerspective}
U.~Zerbst, R.~A. Ainsworth, H.~T. Beier, H.~Pisarski, Z.~L. Zhang, K.~Nikbin,
  T.~Nitschke-Pagel, S.~M{\"{u}}nstermann, P.~Kucharczyk, D.~Klingbeil, {Review
  on fracture and crack propagation in weldments - A fracture mechanics
  perspective}, Engineering Fracture Mechanics 132 (2014) 200--276.

\bibitem{Zafra2021FractureMicromechanisms}
A.~Zafra, G.~{\'{A}}lvarez, J.~Belzunce, J.~M. Alegre, C.~Rodr{\'{i}}guez,
  {Fracture toughness of coarse-grain heat affected zone of quenched and
  tempered CrMo steels with internal hydrogen: Fracture micromechanisms},
  Engineering Fracture Mechanics 241 (1 2021).

\bibitem{Li2011InfluenceZones}
C.~Li, Y.~Wang, Y.~Chen, {Influence of peak temperature during in-service
  welding of API X70 pipeline steels on microstructure and fracture energy of
  the reheated coarse-grain heat-affected zones}, Journal of Materials Science
  46~(19) (2011) 6424--6431.

\bibitem{BSISO2018BSSpecimens}
{BS ISO}, {BS 8571:2018 - Method of test for determination of fracture
  toughness in metallic materials using single edge notched tension (SENT)
  specimens} (2018).

\bibitem{E1820}
{ASTM International}, {ASTM E1820-23b: Standard Test Method for Measurement of
  Fracture Toughness} (2023).

\bibitem{Cupertino-Malheiros2024OnSusceptibility}
L.~Cupertino-Malheiros, T.~K. Mandal, F.~Th{\'{e}}bault,
  E.~Mart{\'{i}}nez-Pa{\~{n}}eda, {On the suitability of single-edge notch
  tension (SENT) testing for assessing hydrogen-assisted cracking
  susceptibility}, Engineering Failure Analysis 162 (8 2024).

\bibitem{Verstraete2013DeterminationMeasurements}
M.~A. Verstraete, R.~M. Denys, K.~Van~Minnebruggen, S.~Hertel{\'{e}},
  W.~De~Waele, {Determination of CTOD resistance curves in side-grooved
  Single-Edge Notched Tensile specimens using full field deformation
  measurements}, Engineering Fracture Mechanics 110 (2013) 12--22.

\bibitem{Hioe2017ComparingTesting}
Y.~Hioe, S.~Kalyanam, G.~Wilkowski, {Comparing unloading compliance and d-c EP
  technique during SENT testing}, International Journal of Pressure Vessels and
  Piping 156 (2017) 79--88.

\bibitem{Harris2022}
Z.~D. Harris, J.~T. Burns, Evaluation of strain-induced artifacts in crack
  length measurements via the direct current potential difference technique,
  in: J.~Kang, P.~C. McKeighan, G.~Dahlberg, R.~Kemmerer (Eds.), Evaluation of
  Existing and New Sensor Technologies for Fatigue, Fracture, and Mechanical
  Testing, ASTM International, West Conshohocken, PA, 2022, pp. 138--159.

\bibitem{VanMinnebruggen2017CrackMeasurement}
K.~Van~Minnebruggen, S.~Hertel{\'{e}}, M.~A. Verstraete, W.~De~Waele, {Crack
  growth characterization in single-edge notched tension testing by means of
  direct current potential drop measurement}, International Journal of Pressure
  Vessels and Piping 156 (2017) 68--78.

\bibitem{Somerday2013ElucidatingConcentrations}
B.~P. Somerday, P.~Sofronis, K.~A. Nibur, C.~San~Marchi, R.~Kirchheim,
  {Elucidating the variables affecting accelerated fatigue crack growth of
  steels in hydrogen gas with low oxygen concentrations}, Acta Materialia 61
  (2013) 6153--6170.

\bibitem{Cochrane2013HAZSteels}
R.~C. Cochrane, {HAZ Microstructure and Properties of Pipeline Steels}, in:
  Proceedings of the International Seminar on Welding on High Strength Pipeline
  Steels, CBMM and The Minerals, Metals and Materials Society, USA, 2013, pp.
  153--188.

\bibitem{Nolan2005HardnessPipeline}
D.~Nolan, Z.~Sterjovski, D.~Dunne, {Hardness prediction models based on HAZ
  simulation for in-service welded pipeline}, Science and Technology of Welding
  and Joining 10~(6) (2005) 681--694.

\bibitem{Hart1991CompositionalSteels}
P.~H. Hart, P.~L. Harrison, {Compositional parameters for HAZ cracking and
  hardening in C-Mn steels}, Welding International 5~(7) (1991) 521--536.

\bibitem{Chatzidouros2019FractureConditions}
E.~V. Chatzidouros, A.~Traidia, R.~S. Devarapalli, D.~I. Pantelis, T.~A.
  Steriotis, M.~Jouiad, {Fracture toughness properties of HIC susceptible
  carbon steels in sour service conditions}, International Journal of Hydrogen
  Energy 44~(39) (2019) 22050--22063.

\bibitem{Asme2023HydrogenB31.12-2023}
{ASME}, {Hydrogen Piping and Pipelines: ASME Code for Pressure Piping,
  B31.12-2023}, Standard ASME B31.12-2023, American Society of Mechanical
  Engineers (2023).

\bibitem{SanMarchi2010FractureHydrogen}
C.~San~Marchi, B.~P. Somerday, K.~A. Nibur, D.~G. Stalheim, T.~Boggess,
  S.~Jansto, {Fracture and fatigue of commercial grade API pipeline steels in
  gaseous hydrogen}, in: Proceedings of the ASME 2010 Pressure Vessels \&
  Piping Division / K-PVP Conference PVP2010, Bellevue, Washington, USA, 2010,
  pp. 2010--25825.

\end{thebibliography}
\end{document}